\documentclass[acmsmall]{acmart}
\renewcommand\footnotetextcopyrightpermission[1]{} % removes footnote with conference information in first column
\usepackage{fancyhdr}

\usepackage{url}  %Required
\usepackage{graphicx}  %Required
\usepackage{enumerate} 

\usepackage{booktabs} % For formal tables

\usepackage{amsbsy}
\usepackage{multirow}
\usepackage{multicol}
\usepackage[caption = false]{subfig}
\usepackage{graphicx}
\usepackage{color}
\usepackage{amsmath}
\usepackage{xcolor}
\usepackage{soul}
\usepackage{algorithm}
\usepackage{algorithmic}

\usepackage{array}
\newcolumntype{L}[1]{>{\raggedright\let\newline\\\arraybackslash\hspace{0pt}}m{#1}}
\newcolumntype{C}[1]{>{\centering\let\newline\\\arraybackslash\hspace{0pt}}m{#1}}
\newcolumntype{R}[1]{>{\raggedleft\let\newline\\\arraybackslash\hspace{0pt}}m{#1}}

\newcolumntype{X}[1]{>{\raggedleft\let\newline\\\arraybackslash\hspace{0pt}}p{#1}}
\newcolumntype{Y}[1]{>{\raggedright\let\newline\\\arraybackslash\hspace{0pt}}p{#1}}

\usepackage{amsthm}

\usepackage{breqn}
\AtBeginDocument{%
  }

\setcopyright{acmcopyright}
\copyrightyear{2018}
\acmYear{2018}
\acmDOI{XXXXXXX.XXXXXXX}

\acmJournal{JACM}
\acmVolume{37}
\acmNumber{4}
\acmArticle{111}
\acmMonth{8}

\begin{document}

%%
%% The "title" command has an optional parameter,
%% allowing the author to define a "short title" to be used in page headers.
\title{Towards Trustworthy AI-Empowered Real-Time Bidding for Online Advertisement Auctioning}

%%
%% The "author" command and its associated commands are used to define
%% the authors and their affiliations.
%% Of note is the shared affiliation of the first two authors, and the
%% "authornote" and "authornotemark" commands
%% used to denote shared contribution to the research.
\author{Xiaoli Tang}
% \authornote{Both authors contributed equally to this research.}
\email{xiaoli001@ntu.edu.sg}

\author{Han Yu}
% \authornotemark[1]
\email{han.yu@ntu.edu.sg}
\affiliation{%
  \institution{Nanyang Technological University}
  \streetaddress{50 Nanyang Avenue}
  \country{Singapore}
  \postcode{639798}
}

% \author{Lars Th{\o}rv{\"a}ld}
% \affiliation{%
%   \institution{The Th{\o}rv{\"a}ld Group}
%   \streetaddress{1 Th{\o}rv{\"a}ld Circle}
%   \city{Hekla}
%   \country{Iceland}}
% \email{larst@affiliation.org}

% \author{Valerie B\'eranger}
% \affiliation{%
%   \institution{Inria Paris-Rocquencourt}
%   \city{Rocquencourt}
%   \country{France}
% }

% \author{Aparna Patel}
% \affiliation{%
%  \institution{Rajiv Gandhi University}
%  \streetaddress{Rono-Hills}
%  \city{Doimukh}
%  \state{Arunachal Pradesh}
%  \country{India}}

% \author{Huifen Chan}
% \affiliation{%
%   \institution{Tsinghua University}
%   \streetaddress{30 Shuangqing Rd}
%   \city{Haidian Qu}
%   \state{Beijing Shi}
%   \country{China}}

% \author{Charles Palmer}
% \affiliation{%
%   \institution{Palmer Research Laboratories}
%   \streetaddress{8600 Datapoint Drive}
%   \city{San Antonio}
%   \state{Texas}
%   \country{USA}
%   \postcode{78229}}
% \email{cpalmer@prl.com}

% \author{John Smith}
% \affiliation{%
%   \institution{The Th{\o}rv{\"a}ld Group}
%   \streetaddress{1 Th{\o}rv{\"a}ld Circle}
%   \city{Hekla}
%   \country{Iceland}}
% \email{jsmith@affiliation.org}

% \author{Julius P. Kumquat}
% \affiliation{%
%   \institution{The Kumquat Consortium}
%   \city{New York}
%   \country{USA}}
% \email{jpkumquat@consortium.net}

%%
%% By default, the full list of authors will be used in the page
%% headers. Often, this list is too long, and will overlap
%% other information printed in the page headers. This command allows
%% the author to define a more concise list
%% of authors' names for this purpose.
\renewcommand{\shortauthors}{Tang and Yu}

%%
%% The abstract is a short summary of the work to be presented in the
%% article.
\begin{abstract}
Artificial intelligence-empowred Real-Time Bidding (AIRTB) is regarded as one of the most enabling technologies for online advertising. It has attracted significant research attention from diverse fields such as pattern recognition, game theory and mechanism design. Despite of its remarkable development and deployment, the AIRTB system can sometimes harm the interest of its participants (e.g., depleting the advertisers' budget with various kinds of fraud). As such, building trustworthy AIRTB auctioning systems has emerged as an important direction of research in this field in recent years. Due to the highly interdisciplinary nature of this field and a lack of a comprehensive survey, it is a challenge for researchers to enter this field and contribute towards building trustworthy AIRTB technologies.
This paper bridges this important gap in trustworthy AIRTB literature. We start by analysing the key concerns of various AIRTB stakeholders and identify three main dimensions of trust building in AIRTB, namely security, robustness and fairness. For each of these dimensions, we propose a unique taxonomy of the state of the art, trace the root causes of possible breakdown of trust, and discuss the necessity of the given dimension. This is followed by a comprehensive review of existing strategies for fulfilling the requirements of each trust dimension. In addition, we discuss the promising future directions of research essential towards building trustworthy AIRTB systems to benefit the field of online advertising. 
%To the best of our knowledge, this survey is the first-of-its-kind literature targeted at trustworthy RTB auctioning system. 
\end{abstract}

%%
%% The code below is generated by the tool at http://dl.acm.org/ccs.cfm.
%% Please copy and paste the code instead of the example below.
%%


%%
%% Keywords. The author(s) should pick words that accurately describe
%% the work being presented. Separate the keywords with commas.

%%
%% This command processes the author and affiliation and title
%% information and builds the first part of the formatted document.
\maketitle

\section{Introduction}
Recent years have witnessed widespread adoption of online advertising, which has become the dominant sector in the advertising industry. Compared with traditional television, radio, newspaper, magazines and billboards, online advertising not only provides advertisers with an alternative option to diversity their strategies to reach more potential customers via the Internet, but also allows them to personalize ads to viewers in a real-time and cost-effective manner \cite{wang2017display}. The key enabling technology for online advertising is Real-Time Bidding (RTB), which refers to the algorithmic trading of online advertising opportunities (a.k.a. ad impressions) through artificial intelligence (AI)-empowered real-time auctioning \cite{wang2017display}. In RTB, the entire auction process for each impression usually takes less than 100 milliseconds before the ad is positioned. By automating the auctioning process involving a large number of available inventories among ad publishers, RTB has significantly transformed the online advertising marketplace.

Compared with other types of online advertising, RTB offers a more streamlined, efficient and targeted purchasing process for advertisers. It promotes user behavior targeting based on user data rather than contextual data, and focuses on the most relevant inventory which can result in high returns on investment for the advertisers. Currently, there have been two surveys on the topic of RTB \cite{wang2017display, liu2020research}. They review RTB from the perspective of algorithm design with the aim of maximizing the revenue or other key performance indicators for different participants of the ad delivery process. 
Despite its development, RTB still faces challenges which threaten its trustworthiness. 
Firstly, RTB systems face many security threats. On the one hand, there have been diverse types of frauds which can deplete advertisers' budgets. 
On the other hand, ads can be injected into the publishers' pages by malicious participants, which brings no revenue to the publishers or even damage their reputation. 
% Moreover, existing RTB algorithms may cause privacy leakage, especially user privacy leakage. Take it for example, users' privacy may be violated without their awareness by the widespread malvertising, which refers to propagating malware through online ads. 
Secondly, RTB systems face challenges that demand high levels of algorithmic robustness. As shown in \cite{wang2017display, yang2021multi}, RTB suffers from the sample selection biases (SSBs), which characterize the systematic distinction  between the data distributions in the training space and the inference space. 
Last but not least, RTB systems face users from diverse demographic backgrounds. Thus, fair treatment of the users is an important consideration \cite{datta2014automated}. 

As trustworthy AI research starts to gain traction in recent years, works on enhancing the trustworthiness of RTB systems have also emerged. Nevertheless, there is currently no comprehensive survey on trustworthy AI-empowered RTB (AIRTB) techniques. In this paper, we attempt to bridge this gap. This paper contributes to the trustworthy AI literature in the following ways:
\begin{enumerate}
    \item We provide a detailed analysis of the RTB technology ecosystem, with focus on the diverse stakeholders involved and the trustworthy AI dimensions important to them.
    \item We propose a unique multi-tiered taxonomy of trustworthy AIRTB based on the major techniques supporting AIRTB, and summarize trustworthy AI related challenges in each part. In lower tiers of this taxonomy, we summarize the key techniques supporting the security, robustness and fairness aspects of AIRTB. To the best of our knowledge, it is the first such taxonomy on this topic, and provides new perspectives to existing works in this field.
    \item We discuss the main metrics adopted in existing approaches to experimentally evaluate the performance of trustworthy AIRTB approaches, thereby, providing readers with a useful guide on experiment design.
    \item We outline promising future research directions towards building trustworthy AI-empowered real-time auctioning systems for online advertising. For each direction, we analyse the limitations in the current literature and propose potential ways forward.
\end{enumerate}
Through this survey, we aim to provide researchers and practitioners with an informative overview of trustworthy AI-empowered real-time auctioning to help them enter this interdisciplinary field.

\section{An Overview of Trustworthy AIRTB}
\label{sec:rtba}
This section provides backgrounds on the topic of Trustworthy AIRTB \cite{liu2020research}. We start by introducing the main terminologies. Then, we illustrate the AIRTB ecosystem and the commonly adopted revenue models. Lastly, we summarize the trust building requirements by various AIRTB stakeholders.  

\begin{table*}[ht]
  \centering
  \caption{Stakeholders of a typical AIRTB ecosystem.}
  \resizebox*{1\textwidth}{!}{
  \begin{tabular}{p{2cm} | p{14cm}}
    \hline
    Stakeholder & Description\\\hline
    %  Users & Potential customer or purchaser of the advertised services or items. A little quantity of anonymous data, or non-personally identifiable data, is transmitted between the browser or the application and the publisher in order to assist the publisher in recognizing which device has connected to its website using a session cookie.\\\hline
    Publishers & Somebody or a certain organization that distributes content in any format for consuming either for free or for a price. \\\hline
    Ad Networks & Institutions that give publishers the possibility to outsource their sales and give media buyers a way to combine inventory and audiences from several sources into a unified purchasing opportunity. They may offer certain techniques, such as distinctive targeting options, creative creation, and optimization, to increase value for both publishers and advertisers. The business models and techniques of Ad Networks may contain elements that are comparable to those that Ad Exchanges provide. Ad networks typically rely on CDNs to transmit advertising-related content. \\\hline
    Supply Side Platforms (SSPs) & Institutions that help publishers run their ad networks and outsource their media sales, from which Demand Side Platforms and Ad Networks buy. They have comparable business models and procedures to ad networks while differing from ad networks in that they do not offer services to Advertisers.\\\hline
    Ad Exchanges & Institutions that offer Publishers and Ad Networks the distribution channels and offer Advertisers the combined inventory, providing a technological platform that enables real-time automated bidding and purchasing based on auctions. The business models and techniques of Ad Exchanges may contain elements that are comparable to those that Ad Networks provide.\\\hline
    Demand Side Platforms (DSPs) &  Institutions that offer integrated and centralized media purchasing from a variety of sources, i.e., Ad Exchanges, Ad networks, and SSPs, generally incorporating real-time bidding capabilities of those sources.\\\hline
    % Agencies & Institutions that organize marketing and advertising campaigns, create and publish advertisements, as well as position media advertisements in the name of its clients. Third-party technique such as ad servers is frequently used by agencies, and they may position advertisements with publishers, ad networks, and other industry participants.\\\hline
    Advertisers & A person, institution, or industry that exploits promotions to advertise a particular good, service, or event in an effort to draw in possible new or recurring customers.\\\hline
    % Data Exchanges & Institutions, or industries that supply user data to DSPs, SSPs, Ad Exchanges, Ad Networks, and Agencies to help them operate more efficiently. I.e., Data Management Platforms (DMP). \\\hline
  \end{tabular}
  }
\label{tab:definition_of_players}
\end{table*}

\subsection{Terminologies}
Table \ref{tab:definition_of_players} lists the key stakeholders (a.k.a. participants\footnote{In this paper, we use ``participants'' and ``stakeholders'' interchangeably.}) in a typical AIRTB ecosystem and their corresponding descriptions according to the Interactive Advertising Bureau (IAB)\footnote{https://wiki.iab.com}.
%which develops commercial norms, offers legal help for the online advertising industry and carries out research, and is regarded as the authority in charge of standardizing online advertising \cite{wiki2014, cai2020threats}. 
An AIRTB ecosystem includes three main types of stakeholders: 1) advertisers, 2) ad networks, and 3) publishers. 
%However, with dozens of ad networks readily accessible via the Internet, both advertisers and publishers were having trouble joining the online advertising market. 
Advertisers need to continually adjust the design of their ad campaigns based on analysis from various ad networks to increase their impact. To maximize revenues, publishers need to selectively subscribe to a number of ad networks based on cost-benefit analysis. Ad networks form Ad Exchanges to offer combined marketplaces for advertisers and publishers to join. 

%Then, advertisers could establish campaigns, set desired targeting, and summarize the data once, while publishers could set up with ad exchanges and generate profit as much as possible without any human intervention \cite{wang2017display}. 

To make it more efficient for advertisers to choose target users and publishers to place their ads, demand side platforms (DSPs) emerge. They are autonomous agents working on behalf of advertisers in an ad exchange. By combining demands, DSPs can enhance the effectiveness and selectivity for advertisers. Similarly, supply side platforms (SSPs) work as the agents to help publishers in an ad exchange sell impressions and optimally manage their inventories  \cite{estrada2017online}. 
%It is worth noting that the distinction between these platforms is fading because of the prospects for financial gain in such businesses. 
Since ad networks, ad exchanges, DSPs and SSPs function together to enable AIRTB services, we refer to them collectively as ``Ad Exchange Networks''. 

%Apart from these participants, there are some auxiliary entities, models and concepts, which are introduced to help the ecosystem. Take it for example, one special kind of Web servers, named Ad servers, are usually adopted to host the content of ads and make such content available to digital platforms. Ad campaigns refer to how much the advertisers have to pay if there ads are shown. The three forms of ad campaigns used by the RTB ecosystem are introduced in Sec. \ref{sec:revenue_model}. 
% Apart from these participants, there usually exists a data exchange, a.k.a. data management platform (DMP) in the ecosystem. DMP collects and stores data of Internet users to profile their preferences and then sells such data to various platforms to help them make automated and customized decisions and adjust campaign parameters accordingly. 

\subsection{Workflow}
\begin{figure}[ht]
\centering
\includegraphics[width=0.8\columnwidth]{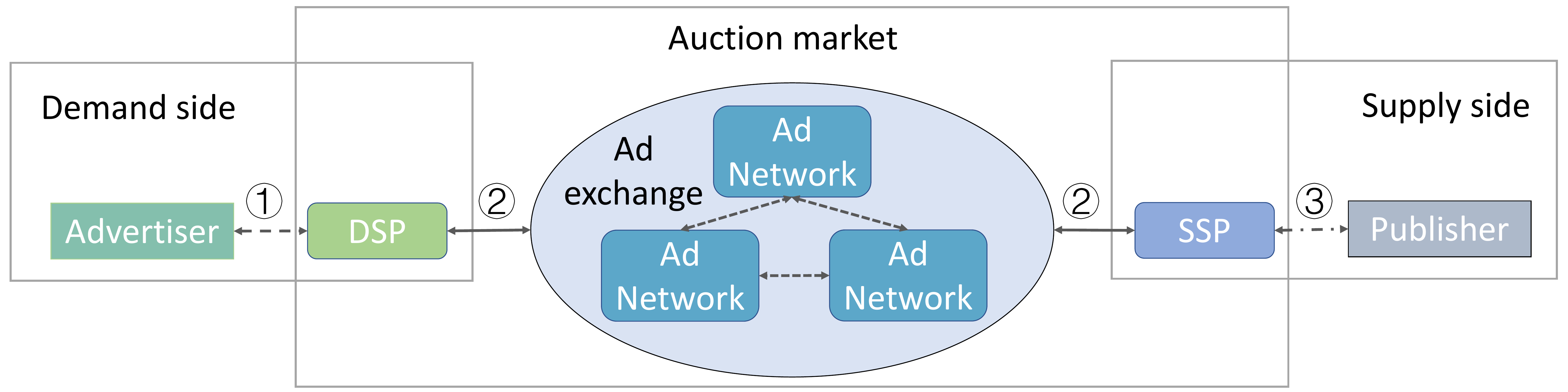}
\caption{The high level overview of a typical AIRTB system. \textcircled{1} The advertiser sets up ad campaigns. \textcircled{2} The auction market strikes the balance between demand and supply by trading campaigns and impressions. \textcircled{3} The publisher notifies the auction market of the impressions. 
% The arc line refers to the requirements posed by the corresponding stakeholders (i.e., the nearest stakeholders) on the whole ecosystem while the arc line with arrow referring to the requirements posed by the corresponding stakeholders on the stakeholder the arrow is pointing to.
}
\label{fig:system_model}
\end{figure}

The high-level overview of a typical AIRTB ecosystem is shown in Figure \ref{fig:system_model}. The advertisers set up ad campaigns in the auction market, while the publishers register ad impressions with the auction market. The auction market execute auctions to trade impressions and ad campaigns in order to strike a balance between demand and supply. 

The typical workflow of each auction is triggered by the emergence of an ad request from a user (e.g., when a user opens a webpage). Upon receiving such ad request, the ad exchange packages it with user information (e.g., the URL of the webpage which the user is visiting, the IP address of the user, etc.) as well as the publisher information (e.g., the context), and transits the packaged bid request to DSPs which might be interested in displaying their advertisers' ads to this user to solicit bids. Based on this bid request, each DSP assesses the potential revenue and possible cost to determine a bid price according to its bidding function, and sends it back as the bid response. Afterwards, the ad exchange selects the winner and determines the market price according to the adopted auction mechanism (e.g., the second-price auction, first-price auction). It then charges the winner and informs the publisher to display the winning ad. 

To improve the efficiency of ad delivery, the advertisers and the ad exchange networks often adopt the Online Behavioural Advertising (OBA) technology, which collects users’ actions, constructs models to estimate user preferences, and thereafter shows ads tailored to the estimated preferences. To achieve this, they collect and share information about both the users and the publishers by utilizing browser cookies. 

% \subsection{Ad Delivery Process in RTB}
% \begin{figure}[t!]
% \centering
% \includegraphics[width=1\columnwidth]{image/ad_delivery_process.pdf}
% \caption{Illustration of the Ad delivery process in RTB in the browser environment.
% }
% \label{fig:ad_delivery_process}
% \end{figure}
% To make it more clear, Fig. \ref{fig:ad_delivery_process} shows the run-time RTB ad delivery procedure in the browser context. The publishers could use a website or a mobile application to deliver ads. The difference between these two channels is the main carrier, the former one using JavaScript snippets  while the later one using Software Developer’s Kit (SDK). 
\subsection{Revenue Models}
\label{sec:revenue_model}
%This subsection goes over how participants of the RTB ecosystem make money. 
Generally, publishers allow advertisers to post their ads on their websites in exchange for commissions from the activities users take. The quantity of impressions, clicks, and user actions constitute three typical basis for the common models that the publishers adopt to generate revenues.
\begin{enumerate}
    \item \emph{Cost Per Impression Mile (CPM)}: Under this revenue model, the fees paid by the advertisers are based on the expense per 1,000 views of the corresponding ad. CPM was designed for conventional advertising systems. It is preferred by the publishers because as long as they display the ads, they can receive revenues without the need to consider user actions (e.g., clicks, conversions). 
    \item \emph{Cost Per Click (CPC)}: Under this revenue model, publishers are paid according to the number of clicks by the viewers on the ads shown in their webpages. Compared to CPM, CPC ensures a higher return on investment for the advertisers as clicks by users are strong indicators of potential interest. 
    \item \emph{Cost Per Action (CPA)}: Under this revenue model, advertisers pay publishers based on the number of the predefined actions taken by the users following the clicks (e.g., purchasing the corresponding products, subscribing to the services). CPC  can be regarded as a special case of CPA. Compared to clicks, the predefined actions following clicks are preferred by the advertisers. As such, CPA is more popular with advertisers. 
    Nevertheless, CPA has its limitations. On the one hand, the publishers are less keen on adopting CPA due to the possibility that fraudulent advertisers may under-report the number of predefined actions performed by the users in order to reduce the commission payout. On the other hand, implementing this model is difficult, particularly when dealing with complicated actions. 
\end{enumerate}

\subsection{Desirable Trustworthy AI Dimensions}
Recent ethics guidelines for AI given by the European Union (EU) \cite{smuha2019eu} state that trustworthy AI ecosystems are supposed to adhere to four main ethical principles: explainability, fairness, prevention of harm, and respect for human autonomy. A variety of trustworthy AI dimensions have been proposed by various organizations and researchers based on these four main principles \cite{brundage2020toward,thiebes2021trustworthy}. In this paper, we follow \cite{liu2021trustworthy} and focus on the following five key dimensions in our discussion about trustworthy AIRTB from the perspectives of the key stakeholders: 1) security, 2) robustness, 3) fairness, 4) explainability and 5) accountability. 

Table \ref{tab:requirements} summarizes the dimensions of trustworthy AI concerned by each stakeholders in an AIRTB system. Specifically, to build trust with the advertisers, the AIRTB system needs to make an effort to improve on all five dimensions. To build trust with the publishers, AIRTB needs to fulfill their requirements from the perspectives of security, robustness, fairness and accountability. As far as the ad exchange networks are concerned, if their requirements of security and accountability are fulfilled, they can establish trust with an AIRTB system. 
%To make it clear, Table \ref{tab:requirements} summarizes the detailed requirements posed by different stakeholders on the AIRTB from the perspectives of security, robustness, fairness, explainability and accountability. 

\begin{table*}[ht]
  \centering
  \caption{Detailed trustworthy AI dimensions required by different AIRTB stakeholders. ($*$) denotes that the corresponding requirement has been studied by existing research.}
  \resizebox*{1\columnwidth}{!}{
\begin{tabular}{|p{2.5cm}||p{4cm}|p{4cm}|p{4cm}|}
\hline
& \textbf{Advertisers} & \textbf{Publishers} & \textbf{Ad Exchange Networks} \\\hline\hline
\textbf{Security}  &  ($*$) Protecting ad campaigns from frauds &  ($*$) Enhancing reputation while gaining revenues &  ($*$) Following legal and sustainable business practices \\\hline
% Users & Being safeguarded against threats, e.g., malvertising, and accessing ads without disclosing privacy & & Being neutrally shown ads to by the ecosystem\\\hline
\textbf{Robustness} & ($*$) Estimating the winning rate, utility and minimum winning prices robustly based on potentially biased historical auction records & Setting the reserve prices appropriately based on potentially noisy data & \\\hline
\textbf{Fairness} &  ($*$) Not being discriminated when the ad exchange networks deliver bid requests & Having fair access to opportunities to initiate ad auctions &\\\hline
\textbf{Explainability} & Being informed by the ad exchange networks of the rationale behind their decisions & &\\\hline
\textbf{Accountability} & Being accountable with clear responsibility attribution & Being accountable with clear responsibility attribution & Being accountable with clear responsibility attribution \\\hline
\end{tabular}
  }
\label{tab:requirements}
\end{table*}

Nevertheless, as the research on trustworthy AIRTB is still in a relatively early stage, the majority of existing studies are concentrated on dimensions of security, robustness and fairness (as shown in Table \ref{tab:requirements}). 
Hence, our survey mainly focuses on these three trustworthy AIRTB dimensions. 

\section{Security}
\label{sec:security}
This section reviews threats facing AIRTB systems and methods to secure them. Specifically, we first briefly summarize the most common attacks targeting AIRTB, and the tools and techniques may used to carry out the attacks. Then, 
% we discuss when and where such attacks can happen. Lastly, 
we discuss and analyse the countermeasures against these attacks. 
% Specifically, this section first briefly summarizes the security requirements of some main stakeholders of the RTB auctioning ecosystem, i.e., users, advertisers, publishers, ad exchange networks. Then, this section moves onto description of the most common threats and attackers may target the ecosystem. Next, we will discuss  when and where such attacks can happen, and what tools and techniques the attackers may use to achieve the objectives. Afterwards, we will sort out and analyse the mitigation strategies and countermeasures used to counter these threats. 

% \subsection{Security}
% As shown in Table \ref{tab:requirements}, security requirements of different  stakeholders vary greatly. As far as the users are concerned, being secured against threats such as malvertising are their urgent needs. 

\subsection{Attackers and Threat Models}
\begin{figure}[ht]
\centering
\includegraphics[width=0.7\columnwidth]{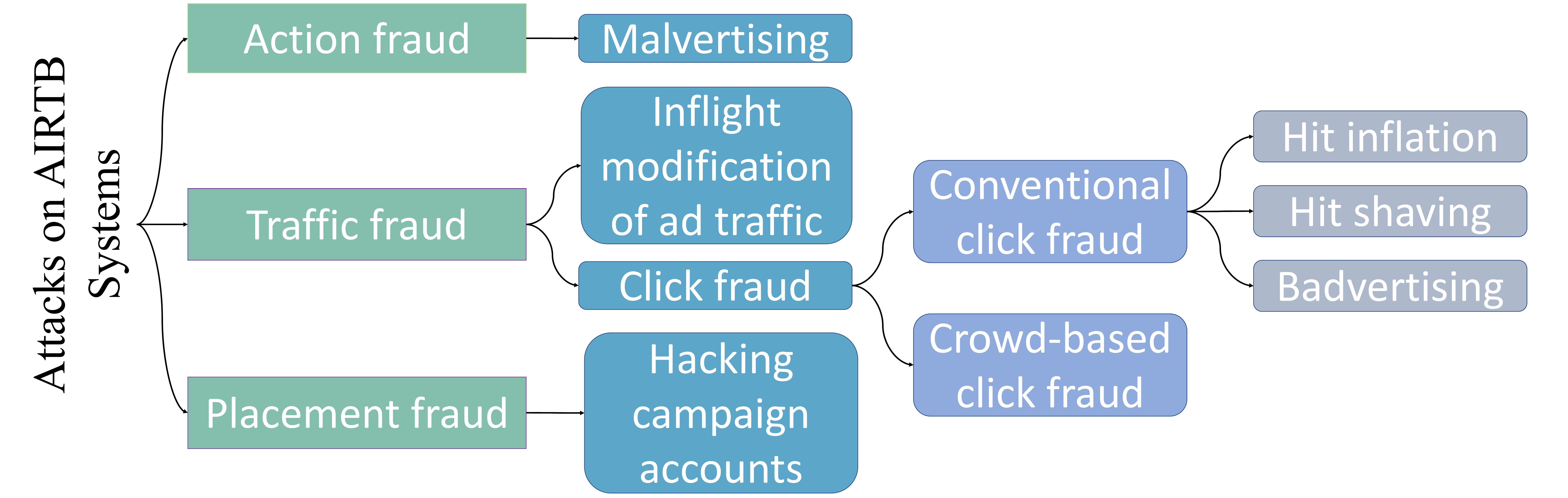}
\caption{Taxonomy of attacks on AIRTB systems.}
\label{fig:attcks}
\end{figure}
The proposed taxonomy of attacks on AIRTB is shown in Fig. \ref{fig:attcks}.
They can be categorized into three major groups: 1) placement fraud, 2) traffic fraud, and 3) action fraud \cite{zhu2017fraud}.  
Placement frauds involve changing or manipulating the information that appears on users' clients or the publisher's websites in order to generate more clicks or impressions. Traffic frauds refer to those that attempt to boost the volume of impressions or clicks from various locations by using fictitious traffics. For instance, malicious attacks can use a crowd or a botnet to artificially boost the number of clicks or impressions on the publishers' websites. Action frauds focus on users' actions to make money. For instance, malicious attackers can hire actors to artificially boost conversion rates with the help of web bots in order to earn additional commissions.

\subsection{Tools and Methods used by Attackers}
Attackers adopt four main types of tools and methods with attacking AIRTB systems: 1) malvertising, 2) inflight modification of ad traffic, 3) click fraud, and 4) hacking campaign accounts. 

\subsubsection{Malvertising}
Malvertising refers to the approach of spreading malwares to vulnerable devices through online advertising \cite{felt2011survey}. 
%In this sense, malvertising can be considered as a practice using online advertising to distribute malwares to vulnerable devices. As described in the last section, the RTB auctioning ecosystem is complicated and includes many participants. 
Due to the complexity of AIRTB which involves numerous redirections among various stakeholders, attackers can insert malicious contents (e.g., malicious ads) into places where the ad exchange networks and publishers failed to anticipate. 
%For instance, it is shown that ad servers can serve JavaScript codes to insert obscured iframe tags into the innocuous websites rather than acquiring truthful ads. The iframe in such process instructs the browsers of the shooter to communicate discreetly with malware servers, enabling the download of PDF exploit files.
Malvertising threats could be launched by AIRTB advertisers as well as publishers. For example, advertisers can easily launch malvertises by inserting malicious ads into legal ad networks. Ad networks may place such ads on the websites of the corresponding publishers, where end users might be misled to click on them. In addition, the publishers can also include malicious contents on their websites, which could inadvertently lead the users to download malwares even without activation process. Flash-based ads constitute one of the widely known types of malvertising \cite{ford2009analyzing}.
%, criminals adopt the Adobe Flash File (i.e., SWF) containing malicious scripts to execute arbitrary orders. The advertisers can reach a wider range of audience by designing SWF files with animated and audio ad creatives, indicating that Flask is open to being employed in malicious attacks. 

\subsubsection{Inflight modification of ad traffic (IMAT)}
In \cite{kshetri2019online}, the authors introduced an ad fraud method known as the Man-In-The-Middle (MITM) attack, which modifies ad traffics in-flight. The Bahama botnet \cite{google2020} is a well-known example of this attack. It enables malwares to force affected devices to provide users with modified ads. Specifically, in the Bahama botnet, the malicious actors manipulate the the Domain Name System (DNS) translations on the compromised devices, and then redirect user traffics to target websites. 
%For instance, an affected user may be discreetly diverted to the malicious actor-controlled servers once he/she click on ads shown on Google or Yahoo. Thus, Yahoo.com or Google.com is translated to IP addresses that adhere to the malicious actors rather that Yahoo or Google. 
In this way, click-through payments are activated by user clicks on the fake ads, resulting in payouts from the advertisers without actual clicks on the intended ads. 
%Generally, Bahama botnet transfers the revenue from larger ad networks to smaller ones and publishers. 
Alternatively, this attack can also be launched through compromised wireless router botnets. This configuration involves turning a malware-infected wireless router into a bot. Then, the botnet master can launch traffic modification attacks in-flight to re-route traffics through such routers. This attack is often performed by public hotspots, which provide free Wifi access to users while inserting ads to boost revenues. 

%The disadvantage of IMAT is that when the users click on the placed ads, the fraudulent ad networks instead of the legitimate ones get the money. Thus, it makes the RTB auctioning ecosystem more vulnerable. In addition, the legitimate advertisers also suffer income loss and reputation loss.  

\subsubsection{Click Fraud}
Click frauds (a.k.a. click spamming, malicious attacks) refer to the automatic or manual attacks that aim to elicit fake clicks on the ads in order to generate illegal revenues \cite{li2014search}. As such attackers click on ads without actually being interested in the contents, they can defraud the advertisers' ad budgets and harm the well-being of the AIRTB ecosystem. Click frauds can be divided into two categories: 1) conventional click frauds, and 2) crowd-based click frauds. 

\textbf{1) Conventional click fraud. } 
Conventional click frauds can be carried out through three main approaches: \textcircled{1} hit inflation, \textcircled{2} hit shaving, and \textcircled{3} badvertising. 

\textcircled{1} \textit{Hit inflation}

This type of click fraud refers to attacks that aim to increase the number of hits to generate revenues for publishers, or on competitors' ads with invalid clicks to deplete their ad budgets \cite{anupam1999security}. It is often performed by publishers and advertisers. 
%And the related two primary kinds of inflation attacks are termed as the publisher click inflation and advertiser competitor clicking. 

\textit{Publisher click inflation. }
This attack takes place when fraudulent publishers purposely inflate the click-through rate without genuine interest in the ads to increase revenue from the ad networks \cite{pooranian2021online}. 
%Since publishers receive payments from advertisers as a result of traffics generated by users, they can make more profits by performing click inflation. However, this makes it easier for dishonest publishers to generate illicit income by raising the volume of user actions, clicks and even impressions on their websites. 
According to the number of fraudsters, publisher click inflation attacks consist of two main types: 1) coalition attacks, and 2) non-coalition attacks \cite{oger2015privacy}. The coalition publisher click inflation attacks are carried out by a group of colluding publishers, while the non-coalition attacks involve only one publisher. Coalition attacks has two main advantages. On the one hand, it is more challenging for countermeasures to detect the relationships among malicious devices and malicious publishers.
%such as the connections among the IP addresses and cookie IDs of the traffic origins and the websites of the malicious publishers, as those malicious publishers are no longer required to reuse the resources to launch additional attacks. 
On the other hand, sharing resources instead of adding more physical resources among malicious devices and malicious publishers lowers the costs of launching attacks. 

\textit{Advertiser clicking inflation. } This attack refers to the case that the malevolent advertisers aim to deplete their competitors' marketing budgets by launching hit inflation attacks \cite{stone2011understanding}. In this way, the attackers can enhance their likelihood to win future ad placement auctions, especially in situations where the daily marketing budgets of the advertisers are limited.

\textcircled{2} \textit{Hit Shaving}

As mentioned in Sec. \ref{sec:rtba}, in contrast to paying the publishers based on the number of clicks, the advertisers generally prefer CPA as they pay publishers based on the desired user actions. Nevertheless, CPA is susceptible to hit shaving attacks \cite{ding2010hybrid}, which are also known as deflation frauds. Through hit shaving attacks, dishonest advertisers decrease the number of clicks from publishers in an imperceptible manner in order to defraud them of the payouts they deserve. 

\textcircled{3} \textit{Badvertising}

Badverting refers to covert click fraud attacks that automatically and silently create clicks on ads once the users access the websites so as to increase the attackers' revenue \cite{gandhi2006badvertisements}.  Compared with traditional attacks based on malwares, badvertiments are more stealthy and generally take the form of phishing attacks and spams. 
%This attack artificially and stealthily increases the number of clicks on ads hosted by the fraudster or unaware associates to generate more revenue for the attacker through advertising. The revenue generated in this way is transferred from the advertiser to the hosting Websites by the fraudster. 
Badvertising consists of two steps: 
(i) delivery, which transmits either corrupt data to users or users to corrupt data; and (ii) execution, which distributes advertisements to the targeted users secretly and automatically. It can be successfully implemented by manipulating the JavaScript codes that the clients' browsers download and run in order to publish advertisements. 
JavaScript snippet files are often inserted into the publishers' Web pages for online advertising systems to function. The JavaScript files will run each time a user accesses the pages and downloads ads from ad servers. When the ads are downloaded, the JavaScript files' frames are updated to include the HTML codes necessary to display the ads. 
The publisher counts how many times the users click on the links to the ad providers' servers using the click-through payment system. 
Consequently, the users are referred to the websites of the ad clients. In order to deploy clicks automatically, Badvertisements run extra malicious scripts. To put it simply, the malicious scripts parse the HTML codes and assemble all links after running and rewriting the frame. Then, they modify the webpages to include the HTML iframes. If the users choose to click the links, the iframes will be triggered in the background and load their contents to take advantage of the users. 
% (i) delivery, which either transfers users to corrupt data or corrupt data to users; and (ii) execution, which automatically and invisibly displays ads to the targeted user. It can be accomplished by corrupting the JavaScript code that is downloaded and executed by the client's browser to publish ads. 
% Online advertisement systems typically work by placing a JavaScript snippet file into a publisher’s Web page. Whenever a user visits this page and downloads an advertisement from the ad server, the JavaScript file will be executed. Downloading the ad causes the frame in the JavaScript file to be rewritten with the HTML code required to show the ad. The publisher relies on the click-through payment process to count the number of times the user clicks on the link to the ad provider’s server. Finally, the user is referred to the ad client’s Website. Badvertisements run extra malicious scripts to automatically deploy clicks. In a nutshell, after running the script and rewriting the frame, the malicious script parses the HTML code and compiles all links. It then changes the Web page to embed an HTML iframe. If the user decides to click the link, the iframe will be activated in the background, and loads its content to exploit the user.

\textbf{2) Crowd-based click fraud. } 
Crowd-based click frauds leverage crowdsourcing \cite{kamar2012combining} to recruit actual individuals to artificially boost ad traffic \cite{tian2015crowd}. 
As crowdsourcing systems are widely available, it is possible to hire a large number of workers to blindly click on a rival advertiser's ads to inflate its expenses. Compared with the conventional frauds mentioned above, crowd-based click frauds possess several characteristics: they often involve a sizable group of people, they generate limited traffic, and the click actions cannot be distinguished from normal click actions. 

%In a nutshell, fraudulent traffic can harm the fame of the publishers, draw reduced number of advertisers, and even subject advertisers to additional fees or fines. 

\subsubsection{Hacking Campaign Accounts}
AIRTB has spawned the tools (e.g., AdWords) which help advertisers launch online campaigns quickly and effectively. However, such tools also make advertisers' accounts vulnerable. Malicious actors can take over the advertisers' accounts and leverage these tools to set up attacks with more significant impact. This attack is referred to as hacking campaign accounts \cite{mladenow2015online}. The campaign accounts may be blocked from legitimate access, or even entered without authorization once they are hacked. 

\subsection{Summary}
\begin{table*}[ht]
  \centering
  \caption{Summary of attacks and the corresponding targeted stakeholders in the AIRTB ecosystem. }
  \resizebox*{1\columnwidth}{!}{
\begin{tabular}{p{2cm}|p{3cm}|p{3cm}|p{0.9cm}|p{0.9cm}|p{0.9cm}|p{1cm}|p{1.4cm}|p{1.4cm}}
\hline
\multirow{2}*{Attack}
& \multirow{2}*{Description} &  \multirow{2}*{Goal} & \multicolumn{3}{c|}{Affected Revenue model} & \multicolumn{3}{c}{Targeted Stakeholder} \\\cline{4-9}
{}  & {} & {} & CPM & CPC & CPA & Ad networks & Publishers & Advertisers \\\hline
Malvertising & Perpetrators injecting malicious codes into the legitimate AIRTB auctioning ecosystem to spread malwares & Generating malicious codes and eventually attempting to redirect users to malicious websites & & & & \checkmark &  \checkmark & \\\hline
Inflight Modification of Ad traffic & Infecting the ecosystem to show modified ads & Generating revenue fraudulently for ad networks and publishers & & \checkmark & &\checkmark &\checkmark & \checkmark \\\hline
Hit inflation & Inflating the real amount of traffic artificially & Gaining financial benefit by exaggerating the number of transactions & \checkmark& \checkmark &\checkmark & &\checkmark & \checkmark\\\hline
Hit shaving & Dishonest advertisers claiming that they receive less traffic than in reality & Dishonest advertisers paying less fees for the traffic & &\checkmark &\checkmark & &\checkmark & \\\hline
Badvertising & Using malicious JavaScript codes to publishing automatic, covert ads& Increasing the number of clicks & &\checkmark & & & &  \checkmark\\\hline
Crowd fraud & Performing malevolent activities by humans against rivals to achieve specific targets & Increasing the fraudulent traffic & \checkmark& \checkmark &\checkmark & & \checkmark& \checkmark\\\hline
Hacking campaign account & Unauthorized access to campaign accounts & Hacker taking control of advertisers' account & & & & & & \checkmark\\\hline
\end{tabular}
  }
\label{tab:attacks_summary}
\end{table*}

\begin{figure}[ht]
\centering
\includegraphics[width=0.85\columnwidth]{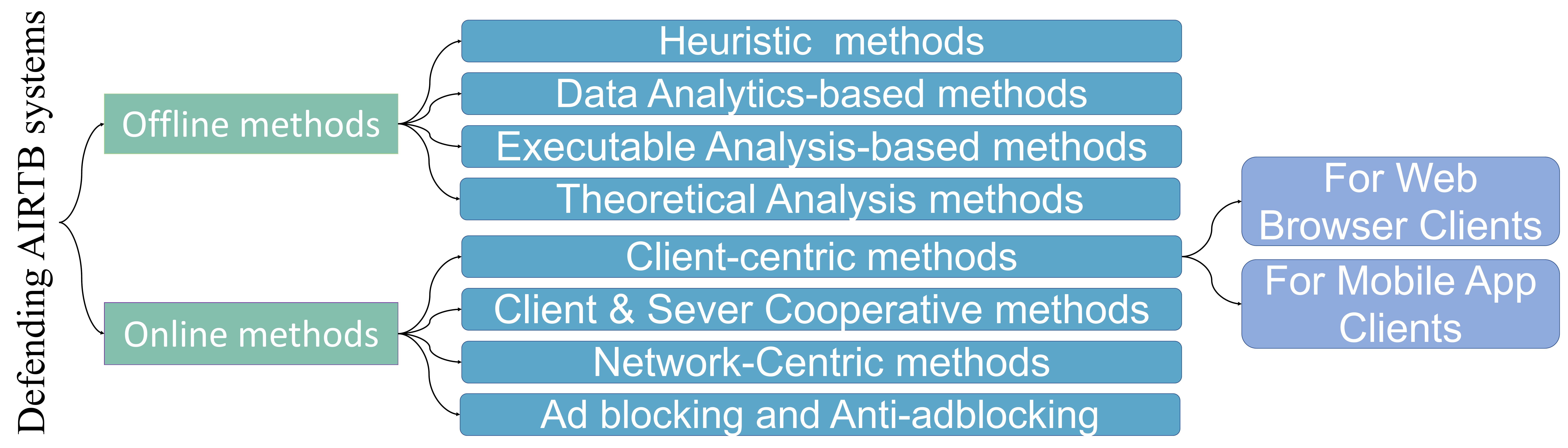}
\caption{Summary of defense methods against attacks on the AIRTB systems.
}
\label{fig:summary_of_defense}
\end{figure}
Table \ref{tab:attacks_summary} summarizes the attacks on AIRTB systems discussed in this section. Among all these attacks, only Inflight Modification of Ad traffic affects all three stakeholders, while hacking campaign accounts, hit shaving, and badvertising only affect one stakeholder. Hacking campaign accounts does not affect specific revenue models. 
Both hit inflation and crowd-based fraud can be applied under all revenue models adopted by AIRTB. Moreover, they both affect publishers and advertisers, while having no effect on ad networks.  
Among the AIRTB stakeholders, publishers and advertisers are the main targets of attacks, while the ad networks are only being targeted by two attacks. CPC is the revenue model that is affected by the most number of attacks, while CPM is affected by the least number of attacks. This is due to the difference in popularity of the revenue models.

\subsection{Defending AIRTB Systems}
After gaining an overview of attacks on AIRTB, we now look into approaches to defending against such attacks. Fig. \ref{fig:summary_of_defense} shows the proposed taxonomy of existing methods against attacks on AIRTB systems.
Based on how and when they take effect, existing AIRTB defense methods can be classified into two main categories: 1) offline methods and 2) online methods. The former are designed to detect and mitigate attacks before and after the ads have been placed. The latter are activated during the ad placement process. In this sense, they complement each other.  

\subsubsection{Offline Methods}
There are four main categories of offline defense methods for AIRTB: 1) Heuristic methods, 2) Data Analytics–based methods, 3) Executable Analysis-based methods, and 4) Theoretical Analysis methods. 

\textbf{1) Heuristic Methods}. 
In the early stage of development for AIRTB, most ad exchange networks clean up malicious ads based on heuristics \cite{mladenow2015online, tian2015crowd}. For example, if an advertiser finds that it received many clicks from the same IP address without actual purchase, it can exclude future clicks from this IP address (or even the entire region in which the IP address is located). Such methods generally incur high labor costs and tend to become ineffective quickly due to the rapid evolution of attackers' strategies. 

\textbf{2) Data Analytics–based Methods}. 
As machine learnig-based data analytics techniques develop in recent years, researchers have started to leverage them for detecting ad frauds and malicious ads.
Based on anonymized data produced for a data-mining competition in 2012, 
% in which the task is to weed out malicious publishers creating click fraud from a collection of publishers, which includes both malicious ones and normal ones,  
\cite{oentaryo2014detecting} reveals the following discoveries. Firstly, to accurately detect frauds, it is essential to analyse the potential features embedded in fine-grained time-series. 
Secondly, the most effective approaches to nonlinear classification tasks that are strongly unbalanced, combined with heterogeneous variables and noisy or missing patterns are those that combine numerous traditional data-mining techniques. 
However, it is worth noting that as the competition took place prior to the wide adoption of deep learning. Deep learning techniques can improve the ability to detect ad fraud, particularly in terms of feature engineering capacity \cite{badhe2017click}. 
% They coarsely report the use of neural networks in fraud detection, with some detailed information omitted.

Based on the findings that publishers involved in click frauds tend to receive higher return on investment (ROI) than honest ones, the Viceroi defense method has been proposed \cite{dave2013viceroi}. It consists of both the offline and the online components. The offline component exams click logs across a range of periods to filter out fraudulent clicks as well as geographic areas where the distribution of revenue per user are abnormal. The online component determines whether a specific click belongs to the abnormal region, based on the aforementioned considerations. 
%To some extent, the higher ROI, regardless of the kind of techniques employed by the click fraud to perform the fraud, just verifies the expanded danger that the click fraud need to carry. 
In \cite{dave2012measuring}, the authors analysed a preventive method used to identify click frauds by ad exchange networks based on a variety of information from a unique publisher website (e.g., mouse movements), and added fake ads to the website to elicit malicious participants' clicking data. 
Similar to adopting the Bayesian method to handle false positive and false negative cases in data analytics,  \cite{pearce2014characterizing} proposed a Bayesian-based approach to detect fraud from click streams. 

In cases where prior ground truth information is unavailable, click fraud detection is more challenging \cite{berrar2016learning}. 
%To address this problem, the authors of \cite{berrar2016learning} discussed the click fraud estimation method applicable to cases with no knowledge of the ground truth. Specifically, the method in this paper is founded on the premise that the estimations of the ground models do not match the unidentified ground reality and thus the class labels that are created automatically have inherent uncertainty. According to this paper, it is better for supervised learning methods that are based on automatically labeled data to take the advantage of the explanation of the ground model's and new classifier's divergent estimations. 
In \cite{poornachandran2017demalvertising}, ads obtained by using both static and behavior analysis are used to analyse fraudulent behavhours. These ads are first divided into nine features, and then inputted to a Support Vector Machine (SVM) to identify malicious ones inserted by publishers. 
In addition, \cite{li2012knowing} proposed a malvertising detection system - MadTracer - which can be incorporated into ad exchange networks. MadTracer actively crawls information about the AIRTB process and utilizes a decision tree-based method to automatically create a series of fraud detection criteria.

Existing data analytics-based AIRTB defense methods tend to be difficult to deploy. Moreover, as these methods are generally developed based on known attacks, they are unable to detect new attacks. 

\textbf{3) Executable Analysis-based Methods}. 
In AIRTB mobile advertising, third-party libraries are required. However, since ad libraries are granted the same level of authorization as their hosting applications, this can lead to serious security and privacy issues if these libraries are compromised. As shown by \cite{cai2020threats}, security flaws can be found early by analysing the malicious executables utilized in malvertising, mobile advertising libraries, and the mobile applications. 

%Specifically, during an application installation process, if the application gets the permissions authorized by the user, the ad libraries then get the same permissions. To avoid these problems, both mobile applications and ad library static analysis should be given top priority in the process of defense.
% In malvertising, the exploit kit refers to the malicious executable profile which is submitted to the victim’s PC. As such, we can use the static analysis to identity such malicious activities. 

In \cite{grace2012unsafe}, the authors developed a system, which can detect various types of risks in AIRTB (e.g., gathering sensitive user data, retrieving code from the web). Their findings suggest that mobile application stores should impose stricter rules on applications that have embedded ad libraries. 
In \cite{crussell2014madfraud}, the authors developed an analysis tool - MAdFraud - to detect ad fraud by concurrently executing a number of applications in emulators. It performs detection in three steps: 1) constructing HTTP request trees, 2) recognizing ad request pages via machine learning, and 3) detecting clicks in the constructed trees based on the adopted heuristics. 
% This work also presents a study based on more than 130,339 applications that are retrieved from 19 Android markets, such as third-party markets and Play, as well as 35,087 applications that are supplied by a security firm and may carry malwares. In this report, it is shown that 27 out of the selected 130,339 applications create clicks without any user input while approximately 30\% of the selected applications that display ads to users send background ad requests while they are active in the backend. 
In \cite{liu2014decaf}, DECAF was proposed to automatically detect multiple types of ad placement frauds in Windows-based mobile platforms. It focuses on the user interface (UI) state transition graph and exploits automated application navigation and optimizations to scan a large number of visual elements in a short time, and determine whether ads within a certain application are in violation of predefined standards governing ad placement and presentation. In addition, to study the malware executables adopted in malvertising, \cite{sakib2015automated} proposed an automated framework which can simulate suspicious browsing activities based on over 800 real-world malicious executables.

The majority of the approaches in this group are susceptible to  attacks such as click-farms and can be rendered ineffective by sophisticated ad frauds (e.g., botnet ad frauds). In addition, the performance of such methods often depends on the tuning of key parameters, which makes them difficult to deploy. 

\textbf{4) Theoretical Analysis Methods}. 
Theoretical analysis of the security of AIRTB can offer useful ideas for building practical solutions. 
%In this subsection, we pay attention to the theoretical investigation part. 
Game theoretic approaches are widely adopted in this area to study the interactions between defenders and attackers as well as other stakeholders in an AIRTB auctioning ecosystem. Many believed that ad networks (e.g., Google) would lose money if it compensated advertisers for fraudulent clicks, which implies that there is no financial motivation for ad networks to combat fraud. However, analysis in \cite{mungamuru2008should} found that ad networks would be worse off in the long term if frauds were unchallenged. 
In \cite{vratonjic2010isps}, a game theoretic model was proposed to study the botnet-driven ad fraud issue. It reveals that, in certain cases, ad networks are unable to resolve the issue of ad fraud on their own and needs to incur additional costs to elicit help from trusted third-parties. 
%However, addressing the issue of ad fraud is a totally different matter, regardless of whether the ad network is financially motivated or not. 
In \cite{dritsoula2014game}, the authors examined a similar issue with a more sophisticated economic model - the Hotelling Competition-based Game-theoretic model - which is capable of taking into account a wider range of variables. 
\cite{huang2017game, huang2018bayesian} leverage game theoretic modelling to cope with malvertising. In these two works, the malvertiser (i.e., the attacker) and the ad network (i.e., the defender) are players of a Bayesian game, since the the ad network only has partial knowledge of whether it is dealing with a normal advertiser or an attacker. 

Although these works shed light on the motivations and trade-offs of attackers and defenders, they generally lack experimental evaluation results to investigate the realism of the findings in practice, making them difficult to apply. 

\subsubsection{Online Methods} 
Online approaches defending AIRTB can be divided into four categories: 1) Client-Centric Methods, 2) Client \& Server Cooperative Methods, 3) Network-Centric Methods, and 4) Adblocking and Anti-Adblocking methods. 

\textbf{1) Client-Centric Methods}. As the name suggests, client-centric methods attempt to address security threats at the client side, either web browsers or mobile apps. Since the execution settings as well as integration techniques for these two types of clients are significantly different, researchers have proposed different methods for them. 

\textcircled{1} \textit{For Web Browser Clients}

Tripwire \cite{reis2008detecting} is thought to be the milestone client-side defense solution for AIRTB in web browser environments to detect HTTP changes made to the web pages. It was considered as a competitor of HTTPS regarding the ad since it is less expensive. 
%These days, with the help of techniques like letsencrypt.org, it is pretty affordable to implement the HTTPS protocol. However, Tripwire still plays an important role in the field as it remains an innovative client-side method for providing fundamental security audits for web servers. 
One drawback of Tripwire is that it cannot cope with a large number of attacks taking place simultaneously at the endpoint. Well-written malicious codes as well as well-written browser extensions can still circumvent Tripwire to manipulate the target webpages. In addition, Tripwire struggles from the lack of a reliable channel for communication. Consequently, adversaries with enough access privileges can remove the code for integrity verification and fake legitimate responses. 

In \cite{vratonjic2010integrity}, the authors proposed a defense mechanism based on authenticated hash-chains. The fundamental operation of this method is the computation of hash values of web pages. It performed well initially on static webpages. Nevertheless, as websites become increasingly interactive, its effectiveness drops. Today, to assess the integrity of webpages and web applications, we must calculate the hash values of the document object model (DOM) tree, which are provided by the corresponding browser. In \cite{xu2014click}, the authors propose various test methods, which can filter out bots. For example, in order to test mouse events, functionality and behavior of browsers, the authors develop JavaScript snippets. These tests are then used to exam the client ad requests. 

AdSentry \cite{dong2011adsentry}, a browser-based defense mechanism, targets ads based on JavaScript to protect website users from attacks such as malvertising. 
It uses a shadow JavaScript engine to entirely mediate the ad script's access to the webpage (including its DOM) without affecting other ad-related functions. This enables flexible regulation of ad script behaviours. 
AdSentry has a policy enforcer that enables end users and the publishers to customize the access permissions for ads. For end users, AdSentry can employ adblockers to automatically detect advertisements and enclose them in a specialized JavaScript wrapper. For the publishers, if the advertisements are encapsulated by unique JavaScript variables, their executions are confined to the shadow JavaScript engine. 

\textcircled{2} \textit{For Mobile App Clients}

Lack of sufficient oversight might enable legitimate advertising service providers to exploit mobile ad libraries to launch attacks on AIRTB from the inside.
To deal with this problem, \cite{pearce2012addroid} proposed AdDroid to separate out privileges related to ad libraries in Android applications. Specifically, AdDroid grants application developers an unique API for advertising. As a result, API requests related to advertising do not receive the same permission as the mobile app itself. 
In \cite{li2015adattester}, a novel verifiable mobile ad framework - AdAttester - was developed based on the ARM TrustZone technique. There are two main types of security primitives included in AdAttester, unforgettable clicks and verifiable displays, both of which are implemented based on the ARM TrustZone hardware root of trust in order to collect proofs that are attached to ad requests for attestation made to ad providers. 
% In work \cite{shekhar2012adsplit}, a technique named AdSplit is developed, considering that malicious software could mimic the actions of advertising libraries. In the proposed AdSplit, the authors have modified the Android operating system. Such modification enables the running of applications and their advertisings as independent processes with distinct user identifiers, frees programs from requesting authorization in the name of the corresponding advertising libraries and offers assistance to verify the veracity of clicks either remotely or locally. Nevertheless, it is unclear whether the Android ecosystem will endorse the idea.
In \cite{iqbal2018protecting}, FCFraud was proposed to target click fraud at the operating system level. It consists of a Linux kernel component which builds HTTP request trees from domain names that are available to the public and associated with advertising. Based on this, it monitors hardware activities such as clicks. If an extrapolated click from the built trees is verified to be not set off by the hardware, the corresponding request is detected as click fraud. AdSherlock \cite{cao2020adsherlock} is a similar approach based on the ad request tree model. In \cite{shi2020clickguard}, the authors proposed a machine learning-based system to distinguish fake clicks from the legitimate ones. It employs a classifier, which is trained using the motion sensor signals from mobile devices. 

Most existing client-centric methods are faced with one main issue: clients are vulnerable to hacking. Once a client is hacked, the security countermeasures within the client are rendered ineffective. As such, defense techniques that utilize both the client and the server have been proposed to get around this limitation. 
%In the following, we will introduce some typical client \& server proposals used to counter threats in the AIRTB systems. 

\textbf{2) Client \& Sever Cooperative Methods}.
In \cite{haddadi2010fighting}, one specific type of baiting ads named bluff ads are developed to be recognizable and clicked by bots or inadequately skilled click farm actors. The clicks on bluff ads are regarded as fraudulent by the server-side component. To make HTTPS more effective in scenarios like caching of Content Delivery Network (CDN), \cite{singh2012practical} proposed a new form of HTTP protocol with integrity, named HTTPi. In order to accomplish this, new modules must be added to web servers as well as web browsers. This approach can reduce ad concerns related to hacking. 
To cope with attacks on the endpoints, \cite{sobel2012methods} proposed a framework for client-server transaction fingerprinting. To verify true clicks (i.e., only those that can be verified as legitimate), \cite{juels2007combating} put forth a new scheme based on requests being verified by client-based cryptography attestations. 
% It is worth noting that the authentication process could make use of web tokens. The fundamental principle is to assign a special identity to each session. 
In \cite{callejo2016independent}, a similar approach was proposed with the goal of improving transparency from the viewpoint of the advertisers. Specifically, it gathers the pertinent data related to each impression (i.e., the URL in which the impression is delivered, the User-Agent which accepts such impression, and interactions between the user and the ad impression like clicks on the ad and mouse movements), and then transmits them to a central server. 
% The timestamp associated with the ad impression and the IP address of the machine which obtains the ad impression are both obtained using the connection made with the server. Finally, the period of the ad impression's exposure will be regarded as the connection's duration. 

In \cite{chenj2014}, the authors envisaged gathering and inputting data related to user mouse movements into machine learning frameworks to aid ad fraud detection. To some extent, this is a useful technique for telling bots and people apart. Nevertheless, its applicability is still limited as currently: 1) the size of user profile data which cover the mouse movements is significantly larger than that of those which do not contain mouse movements; and 2) the data are generally not well-organized. In addition, by adding random noisy information to the automated mouse movements, bots can still trick machine learning algorithms into classifying them as humans. In \cite{aberathne2018smart}, the authors developed a novel framework to identify mobile click bots. It consists of a client-side component and a server-side component. The former is used to gather and screen all events created by mobile clients, while the latter is used to analyse incoming data and filter out bots. 
%Based on the classification model XGBoost and the feature transformation model Cascaded Forests, \cite{thejas2021hybrid} proposed a hybrid machine learning approach, named Cascaded Forest and XGBoost (CFXGB). 

%On the other hand, \cite{benjiaming2015, roozbbehani2018} propose to solve this problem without altering current browsers. In their proposals, a reverse proxy and the protected JavaScript, which possess features including self-defense, anti-tampering, bot-detection, could be utilized to identify security problems in the browser environment. Protected JavaScript makes it possible that attacks can not affect the interactions between web browsers and reverse proxies. 

Current methods in this category tend to focus on the security of only one participant, leaving the other ones still vulnerable to attacks.

\textbf{3) Network-Centric Methods}.
Through investigation of the network edge traffics, network monitoring tools (e.g., intrusion detection systems (IDSs)) can detect fraudulent and malvertising traffics. As such, there have been works taking advantage of IDSs to combat malvertising in AIRTB. 

In malvertising, attackers try to avoid having their web-based exploit kit services banned by hiding them. As a result, more sophisticated malvertising traffic detection methods are required. In order to contaminate client browsers, a web-based attack needs to redirect a web browser to the corresponding landing page, retrieve exploit files, and download infected codes. Based on the design of the web requests, which takes the form of a tree and classifying data based on structural information, \cite{li2012knowing} proposed a method to identify malicious ad traffic. In particular, information retrieval methods are first used to create the index of harmful tree samples. Then, the subtree similarity search algorithm is adopted to identify HTTP flows associated with the malicious exploit kits.  

It is worth noting that these approaches can be expanded and used by large organizations like Internet Service Providers (ISPs) to identify further advertising-related problems, such as ad frauds. However, there are several open challenges to be solved. On the one hand, these approaches need to be made more efficient to handle considerably higher levels of traffic. On the other hand, ISPs might require financial incentives to participate in these approaches. 

\textbf{4) Adblocking and Anti-Adblocking.} 
%The following part will go over methods related to Ad blocking, which possesses the  advantage from the perspectives of security protection and performance improvement, i.e., 
% 1) Firstly,  users are equipped with the permission to block an overwhelming majority of ads shown in adblockers' blocklists, save browsers from carrying out time-consuming ad fetching and rendering processes. 2) Additionally, 
Adblockers can stop multiple types of malvertising. 
% 2) In addition, using adblockers can somewhat help the users preserve their privacy by preventing requests to monitor their online browsing hatis. In this sense, ad blocking can be regarded as a security-enhancing approach. 
Nevertheless, adblocking can be disadvantageous for the publishers whose revenues depend on advertising. To maintain a steady stream of income, the publishers can establish rules which restrict access to the services or contents to people who are prepared to view the ads. Such decisions are based on economic considerations and have been analysed in \cite{ray2017ad} based on game theoretic modelling. In essence, an adblocker functions as a web browser extension and has greater access rights than anti-adblocking scripts, which typically execute within the JavaScript engine sandbox. Thus, adblockers ultimately hold a technical advantage in the rivalry between adblocking and anti-adblocking. Certainly, the rising popularity of adblocking can be perceived as a sign of trouble for the AIRTB ecosystem. 
%It is still unknown if the expenses associated with adblocking is justified. 
However, from a technical standpoint, adblocking and anti-adblocking techniques can help enhance the security of AIRTB systems.

As reported in \cite{nithyanand2016adblocking, iab2016}, plenty of publishers have implemented methods and techniques to identify and disable adblockers. The fundamental approach is through filtering. Firstly, a list is constructed based on feedbacks from the general public. Then, adblockers will block web traffics based on this blacklist. Some web traffics in this blacklist might be equipped with anti-adblockers. To combat these anti-adblockers, some approaches have been developed to improve the existing adblocking methods. In \cite{orr2012approach}, JavaScript source codes undergo static program analysis in order to identify those that display ads. According to \cite{zhu2018measuring}, more than 30\% of the Alexa top 10,000 websites have been equipped with anti-adblockers scripts. The authors of \cite{iqbal2018adgraph} proposed to aggregate information from various sources (including JavaScript, HTML, and HTTP) into a machine learning framework, and adopt AI algorithms to construct a filtering list to detect anti-adblockers. In \cite{iqbal2018adgraph}, the authors developed a scheme to prevent ads from being served via mobile apps. It consists of a machine learning classifier to reject ads using information from packets acquired after intercepting the network interface of the device. 

\subsection{Evaluation Metrics}
Establishing a collection of performance evaluation measures is crucial for the long-term development of security research for AIRTB as it allows for the unbiased comparison of the proposed methods. In this section, we discuss the benchmark evaluation metrics used in the state-of-the-art works. 

\begin{itemize}
    \item \textbf{Accuracy}: This is the most widely used evaluation metric. In current works, accuracy takes various forms, such as the number of recognised malicious attackers \cite{li2015adattester}, precision \cite{chenj2014}, recall \cite{iqbal2018protecting}, the false positive rate \cite{liu2014decaf}, and the false negative rate \cite{xu2014click}. 
    \item \textbf{Efficiency}: Efficiency is another widely adopted evaluation metric. Some works measure the overhead caused by the deployed mitigation strategies, such as processing time \cite{grace2012unsafe}, the overhead on memory and CPU \cite{shekhar2012adsplit}, and the overhead on event throughput (e.g., clicks) \cite{shekhar2012adsplit}. It is worth noting that, efficiency is not used to directly evaluate the effectiveness in terms of combating malicious attacks, but to evaluate deployment complexity. 
\end{itemize}

\subsection{Summary}

Attack detection in AIRTB is an active area of research because malicious actors are constantly coming up with novel attacks. 
Thus, the most important factor in making the entire AIRTB ecosystem secure is through real-time detection of attacks. 
Due to the rapid expansion of AIRTB to meet the demands of the publishers and advertisers, the attack surface of AIRTB also becomes larger. Although numerous efforts are being made to mitigate attacks against AIRTB, the security requirements of all stakeholders still cannot be met by any existing approach in a well integrated manner. Nevertheless, several areas of research are gaining traction: 
\begin{itemize}
    \item Data analytics: Approaches for identifying ad fraud and bots and mitigating malvertising can take advantage of the vast amount of data created in the ads placement process. 
    \item Security improvements: Limiting the permissions of mobile apps and adblocking is an effective solution for improving the security of the AIRTB ad delivery process. 
    \item Theoretical analysis: Game theoretic approaches are frequently adopted to examine issues such as participants' motivations, and resource trade-offs in attacks and defenses.
\end{itemize}

\section{Robustness}
\label{sec:robustness}
\begin{figure}[ht]
\centering
\includegraphics[width=1\columnwidth]{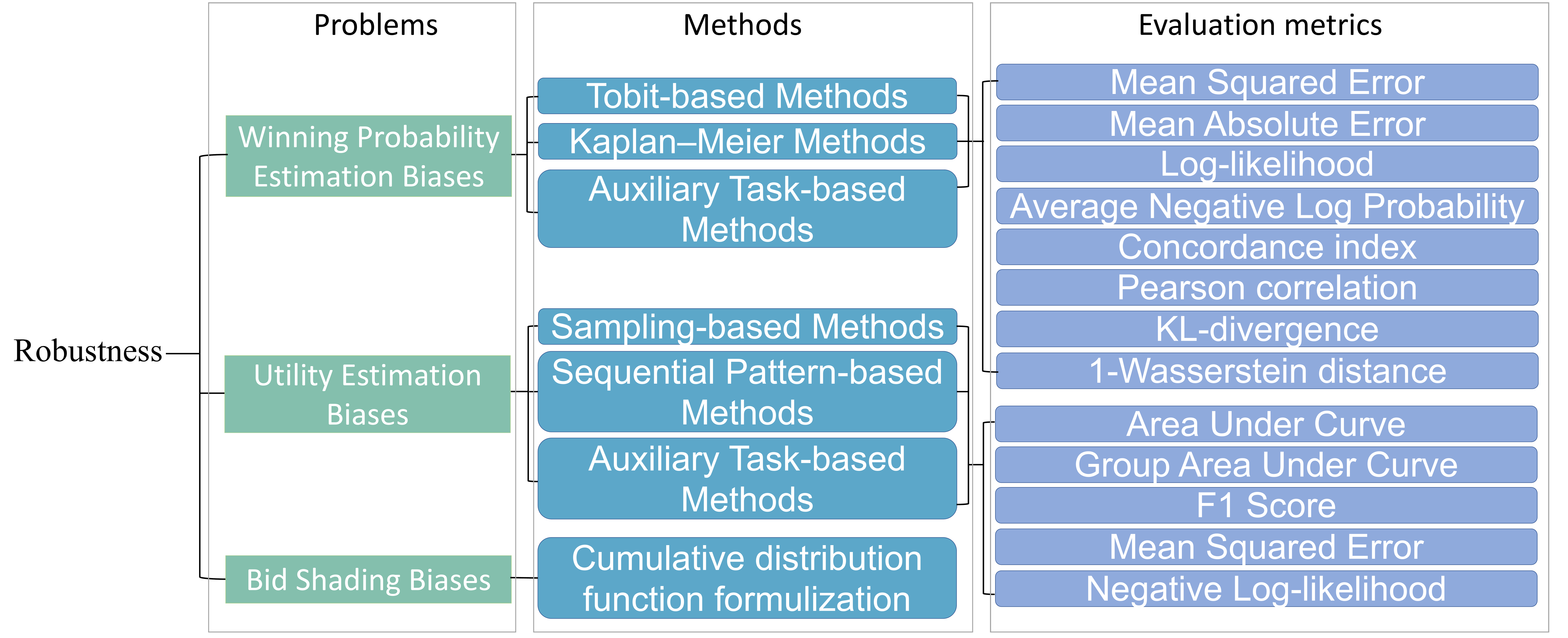}
\caption{Robustness research in AIRTB.
}
\label{fig:robustness_summary}
\end{figure}
Fig. \ref{fig:robustness_summary} provides an overview of the roadmap of research related to robustness in AIRTB. 
Most such works focus on helping AIRTB approaches to be robust against various types of biases. %Therefore, in this paper, we focus on issues related to nonbias. 
According to \cite{liu2021trustworthy}, biases related to AI research can be divided into three categories: 1) discriminatory bias, 2) productive bias, and 3) erroneous bias. 
Discriminatory biases manifest as algorithmic unfairness towards groups or individuals (e.g., the creation of discriminating content or subpar performance for some users) \cite{shah2019predictive}.
Almost all machine learning algorithms are faced with the productive bias problem \cite{hildebrandt2019privacy}.
According to the ``no free lunch theory'' \cite{wolpert1997no}, only predictive models that are biased towards specific distributions can outperform them during modeling. 
%With the help of productive bias, algorithms can tackle specific kinds of issues. 
In general, productive bias is introduced through the presumptions about the learning tasks (via loss function design), the distribution assumption, and the optimization methods. 
Erroneous biases are a specific type of systematic errors caused by unrealistic assumptions. For instance, it is commonly assumed in AI that there is no difference between the distribution of the real data and that of the training data. Nevertheless, due to reasons like sample selection bias \cite{marlin2012collaborative, mehrabi2021survey}, the training data used might not accurately reflect the distribution of the real data. Therefore, if the assumption is incorrect, the learnt model may perform poorly on the test data. In the field of AIRTB, addressing erroneous biases has been the main focus of robustness research. Thus, in the following parts, we use the teams ``bias'' and ``erroneous bias'' interchangeably. 
Current robust AIRTB research focuses on addressing the following types of erroneous biases: 1) winning price estimation biases, 2) sample selection bias (SSB), and 3) bid shading biases in closed (i.e., censored) first price auctions. 

\subsection{Robustness against Winning Probability Estimation Biases}
In second price auction-based AIRTB, when a bid request is received from the ad exchange, DSPs need to calculate a bid price as the bid response to join the corresponding auction. It is important for DSPs to estimate the winning prices in order to optimize their prices \cite{wang2017display}. However, winning price estimation is complicated with the data censorship issue. On the one hand, with second price auction, a DSP only knows the winning price if it wins in the auction. Thus, the winning price observed by each DSP is right-censored: the DSP only knows the winning prices are higher than its bids when it loses the auctions, but not the exact values \cite{wang2017display}. Moreover, in second price auctions, if the reserve price of a specific bid request set by the corresponding SSP is higher than the second highest bid price, but lower than the highest one, the winning DSP needs to pay the reserve price for the bid request instead of the market price. In this sense, the winning price observed by each DSP is left-censored: it only knows the upper bound of the market price. Thus, to get an accurate estimation model for winning probability, these biases caused by the data censorship issue must be addressed. 
%\subsection{Methods used to mitigate the bias in winning probability estimation}
Existing methods for mitigating winning probability estimation biases can be divided into three main categories: 1) Tobit-based methods, 2) Kaplan–Meier methods, and 3) auxiliary task-based methods. 

\subsubsection{Tobit-based Methods} 
This type of methods explicitly divides the loss functions into two parts, one for the winning records with observable winning prices and the other for the losing records with censored winning prices. These two sub-loss functions are then optimized jointly. Wu et al. \cite{wu2015predicting} first proposed to incorporate the censored regression model \cite{greene2005censored} into AIRTB to estimate the winning rate for the target DSP on a given bid price. In this work, linear regression is leveraged for bid prices with observable winning prices, and censored regression is leveraged for bid prices with censored winning prices. The maximum likelihood procedure is used to predict the winning probability of a given bid price. 

Nevertheless, \cite{wu2015predicting} assumes that the winning prices follow a normal distribution, which might not always be true in practice. %In addition, it turns to be difficult to link the physical meaning of the winning prices and the normal distribution. 
To address this problem, \cite{zhu2017gamma} replaced the normal distribution assumption with the gamma distribution assumption, and developed a gamma-based and regularization-aware censored linear regression model.
However, this approach led to further discussions that gamma distributions might also not be realistic. As such, \cite{wu2018deep} explored multiple distributions and combined them with the deep leaning models to investigate the corresponding prediction qualities. Ghosh et al. \cite{ghosh2019scalable} relaxed the assumptions on the distribution of winning probability by adopting a mixture density censored network to learn smooth winning price distributions. 

Apart from the disadvantage caused by the strong assumptions, the aforementioned methods are also unable to estimate the possible ad cost for each specific bid price as they can only perform point estimations of market prices. To address this problem, \cite{ghosh2019scalable} leverages the improved Censored Regression. To address the same problem, \cite{ren2019deep} proposed a deep landscape forecasting model taking advantage of recurrent neural networks (RNNs) to model the conditional winning probability with respect to any given bid price flexibly without imposing any assumptions on the underlying distribution. 

\subsubsection{Kaplan–Meier Methods} 
Methods falling into this category leverage the Kaplan–Meier estimation approach to address the censored data problem in winning probability estimation. Kaplan–Meier estimation is a well-know survival analysis approach used for forecasting a patient's chance of survival after a certain treatment \cite{fleming2011counting, goel2010understanding}, which is similar to the winning probability estimation task. 
In \cite{zhang2016bid}, the authors analysed the striking similarities between the survival analysis and AIRTB winning probability estimation, and leveraged the non-parametric Kaplan-Meier Product-Limit approach \cite{kaplan1958nonparametric} to fit the market price distribution. 
However, this method relies on the counting-based statistics of given sample clusters, making it unable to accurately predict the winning probability for individual bid requests. To address this problem, \cite{wang2022kaplan} proposed to learn the Kaplan–Meier estimation for individual bid requests by predicting two probabilities: 1) the probability of losing a given auction at a certain bid price, and 2) the probability of winning a given auction at a specific market price. 
In most cases, the sequential patterns shown in the features over the price space turn out to be important for winning probability estimation. The deep landscape forecasting model in \cite{ren2019deep} combines the survival analysis for dealing with censorship with RNNs to model the sequential patterns for probability distribution forecasting.

\subsubsection{Auxiliary Task-based Methods}
Yang et al. \cite{yang2021multi} address the winning probability estimation problem by combining it with the utility estimation problem into a multi-task learning problem. The resulting approach is capable of solving the AIRTB winning probability estimation problem as a multi-category classification task. 
Despite performance improvement, this model is faced with the problem of tuning hyperparameters, which are used to weight the importance of the winning probability estimation task versus the utility estimation problem.  

\subsubsection{Evaluation Metrics}
The evaluation metrics commonly adopted to measure the robustness of AIRTB winning probability estimation methods are summarized as follows: 
\begin{itemize}
    \item \textbf{Mean Squared Error (MSE) }: In \cite{wu2015predicting, zhu2017gamma, wang2022kaplan}, MSE between the ground truth winning prices and the estimated ones is adopted to evaluate the effectiveness of the proposed methods:
    \begin{equation}
\label{eq:mse}
\begin{aligned}
MSE = \frac{1}{N}\sum_{i=1}^N (y_i - \hat{y}_i)^2, 
 \end{aligned}
\end{equation}
where $N$ denotes the number of the data samples. $y_i$ and $\hat{y}_i$ are the ground truth value and the estimated value of the $i$-th sample, respectively. The smaller the MSE, the better the performance.
    \item \textbf{Mean Absolute Error (MAE) }: In \cite{wu2018deep}, MAE is adopted as one of the evaluation metrics:
        \begin{equation}
\label{eq:mae}
\begin{aligned}
MAE = \frac{1}{N}\sum_{i=1}^N ||y_i - \hat{y}_i||.
 \end{aligned}
\end{equation}
The smaller the MAE, the better the performance.
    \item \textbf{Log-likelihood (LL)}: It is the log of the density function (or the probability):
            \begin{equation}
\label{eq:ll}
\begin{aligned}
LL = \sum_{i=1}^N \log(f_{x_i}(y_i|\theta)),
 \end{aligned}
\end{equation}
    where $x_i$ is the observed data. $f$ denotes the estimation model with parameters $\theta$. It assesses how likely it is to observe the data using the model. The likelihood that the data are generated by the model increases with the log-likelihood. However, compared with MSE or MAE, log-likelihood performs worse when the values are small. 
    \item \textbf{Average Negative Log Probability (ANLP)}: It measures the probabilities of the bid requests used for testing occur alongside the corresponding market prices:
            \begin{equation}
\label{eq:anlp}
\begin{aligned}
ANLP = -\frac{1}{N}\sum_{i=1}^N \log f_{x_i}(y_i|x_i).
 \end{aligned}
\end{equation}
    \item \textbf{Concordance index (C-index)}: C-index assesses how accurately a model can order samples based on their market prices. 
    \item \textbf{Pearson correlation}: It is adopted to measure the correlation between the estimated results and the ground truth values, and is defined as:
                \begin{equation}
\label{eq:cor}
\begin{aligned}
COR(Y, \hat{Y}) = \frac{\sum_{i=1}^N(y_i-\Bar{y})(\hat{y}_i-\bar{\hat{y}})}{\sqrt{\sum_{i=1}^N(y_i-\Bar{y})^2 \sum_{i=1}^N(\hat{y}_i-\bar{\hat{y}})^2}}.
 \end{aligned}
\end{equation}
The larger the correlation, the better the performance.
\item \textbf{KL-divergence (KL)}: KL between the ground truth distribution $q_Y$ and the estimated distribution $q_{\hat{Y}}$ is formulated as:
                \begin{equation}
\label{eq:kl}
\begin{aligned}
KL(q_Y||q_{\hat{Y}}) = - \sum_x q_Y \log \frac{q_{\hat{Y}}}{q_Y}.
 \end{aligned}
\end{equation}
The smaller the distance, the better the performance.
\item \textbf{1-Wasserstein distance (WD)}: WD between the ground truth distribution $q_Y$ and the estimated distribution $q_{\hat{Y}}$ is formulated as:
                \begin{equation}
\label{eq:wd}
\begin{aligned}
WD(q_Y, q_{\hat{Y}}) = \sum_x |q_Y - q_{\hat{Y}}|. 
 \end{aligned}
\end{equation}
The smaller the distance, the better the performance.
\end{itemize}

\subsubsection{Summary}
\label{sec:wps_methods_summary}
Generally, most AIRTB debiasing methods based on Tobit models tend to be parametric ones, and need to make assumptions on the distributions of the winning probability. 
%However, it may turn out that there is no one distribution that works for all datasets and for all cases. 
Non-parametric methods based on Kaplan–Meier estimation learn the winning probability distribution without making any assumptions. Nevertheless, the majority of the approaches falling into this category need to cluster the data in order to leverage heuristics to boost model accuracy. The clustering operation, to some extent, limits the applicability of these methods to dynamic real world AIRTB data. In addition, methods based on Kaplan–Meier are only able to make coarse-grained predictions using data segments. 
The latest trend is to frame the winning probability from the perspective of multi-tasking learning by incorporating other auxiliary tasks. However, the performance of such approaches is still sensitive to hyperparameters, and requires tedious parameter tuning. 

%In terms of technologies that can be leveraged, there are the following streams:
%\begin{itemize}
%    \item Deep learning method: Methods used for mitigating bias in winning probability estimation can fully explore the advantages of deep learning methods, such as RNN, graph neural networks. 
%    \item Multi-task learning: The introduction of auxiliary tasks, such as utility estimation, designing of bidding strategies, can greatly improve the performance of winning probability estimation. 
%\end{itemize}

\begin{figure}[ht]
\centering
\includegraphics[width=0.5\columnwidth]{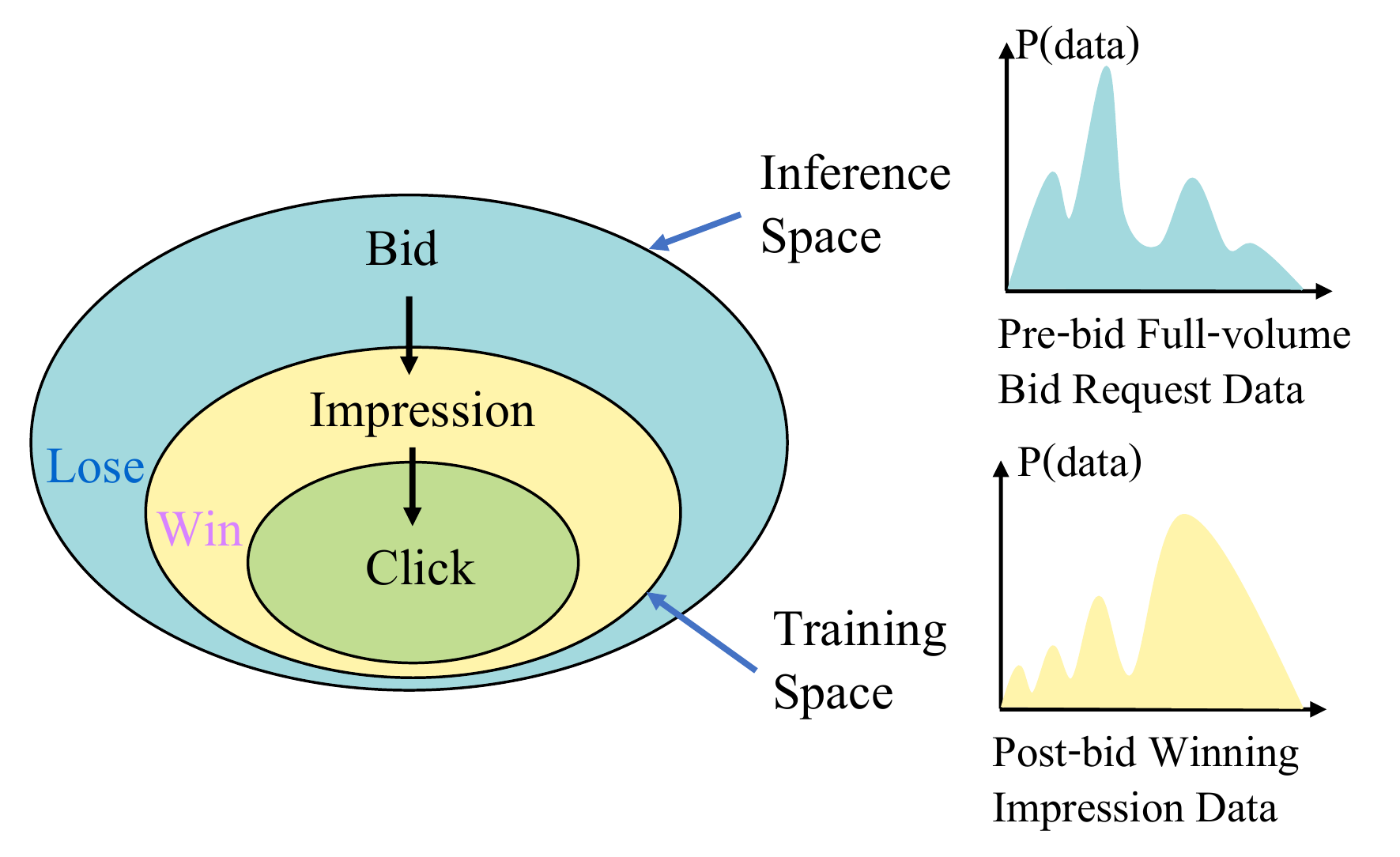}
\caption{Illustration of the sample selection bias (SSB) in RTB auction. The training space only contains winning impression data, whereas the inference space includes all full-volume bid request data \cite{yang2021multi}.
}
\label{fig:cvr_bias}
\end{figure}
\subsection{Robustness against Utility Estimation Biases}
AIRTB is faced with the problem of SSB \cite{zadrozny2004learning}, which refers to the systematic mismatch in data distributions between the training space and the inference space \cite{wen2020entire,yang2021multi}. Existing models for AIRTB utility estimation mostly need labeled data for supervised learning. In practice, the labels on whether the users responded or not (e.g., clicks and conversions) and the market prices for bid requests are only available if the advertiser wins in the corresponding auctions \cite{zhang2016bid, yang2021multi}. As shown in Figure \ref{fig:cvr_bias}, most existing utility estimation approaches are learned only on clicked samples (including click-through rate, i.e., CTR and conversion rate, i.e., CVR), while drawing conclusions about the entire space with all impression samples. In addition, the clicked samples and the converted samples only account for a small fraction of the impression samples. They are biased due to user actions. Thus, the SSB problem greatly reduces the effectiveness of utility estimation models. Compared with CTR, CVR reflects user preferences more strongly (through subscription of service, registration, installation of software, etc.), and is more relevant to advertisers. Thus, in this part, we focus on AIRTB approaches for addressing SSB in CVR prediction.  

%\subsection{Methods used to mitigate the bias in utility estimation}
Existing works for enhancing the robustness of AIRTB against the SSB issue in CVR prediction can be divided into three main categories: 1) Sampling-based Methods, 2) Sequential Pattern-based Methods, and 3) Auxiliary Task-based Methods. 

\subsubsection{Sampling-based Methods} This type of methods attempt to solve the SSB problem in CVR prediction by leveraging sampling approaches. Through the introduction of sampling, the models trained are pulled to fit the true distribution of the entire space instead of just the training space. In particular, \cite{zhang2016bid} addresses the SSB issue in CVR prediction by using rejection sampling to fit the true underlying distribution from observations. However, this method is susceptible to numerical instability when dividing the rejection probability to weight the samples.

\subsubsection{Sequential Pattern-based Methods}
When being shown an ad, the potential responses by the target user form a sequence, which is generally in the form of impression $\to$ click $\to$ conversion \cite{ma2018entire}. Methods used to mitigate SSB in CVR prediction under this category attempt to fully exploit such sequential patterns of user responses. The first such work is \cite{ma2018entire}. It models CVR with all samples via training two auxiliary tasks (i.e., post-view click-through rate and post-view click-through conversion rate). It is extended in \cite{wen2020entire} by modeling two auxiliary actions (i.e., disjoint purchase-related Deterministic Action (DAction) and Other Action (OAction)), which are injected between click and purchase (i.e., conversion). Wang et al. \cite{wang2020delayed} further considered the problem of delayed feedback and proposed ESDF based on neural networks to model the CVR prediction problem from the entire space perspective through combing the benefit of the time delay factor as well as the user sequential behavior pattern. The CTR prediction task is involved in all these studies, which is the task preceding CVR prediction. Intuitively, representations learned from one task may be helpful for the other \cite{o2021analysis}. Therefore, \cite{wei2021autoheri} takes advantage of the interplay of representation learning across multiple tasks to perform neural architecture search to learn the best connections between task-specific layer-wise representations. The proposed AutoHERI frames the CVR prediction in the entire space with automated hierarchical representation integration.

However, all these methods only focus on macro-level behaviors, which help understand subsequent purchase patterns at the granularity of the item level. They overlook the more frequently occuring fine-grained behaviors (e.g., clicks) on detailed elements of items (e.g., user comments, pictures, videos), which are referred to as micro-level behaviors. To some extend, insight into micro-level behaviors can enhance understanding of future macro-level behaviors. As such, \cite{bao2020gmcm} applied Purchase-related Micro-behavior Graph (PMG) to describe the users' micro-level behaviors to transform the CVR prediction problem into a graph classification problem. Work \cite{wen2021hierarchically} extended \cite{ma2018entire} and \cite{wen2020entire} by constructing the complete user sequential behavior graph, where the micro-level and macro-leve behaviors are hierarchically encapsulated as the one-hop and two-hop post-click nodes in a unified framework, respectively. Inspired by studies used to eliminate bias in recommender systems, \cite{zhang2020large} frames the SSB problem in CVR prediction from a causal perspective, and accounts for the causes of missing not at random \cite{little2019statistical}. They proposed two causal estimators for CVR prediction, which adapt the missing not at random mechanism to be trained on a perfect dataset where all exposed items are clicked by users.

\subsubsection{Auxiliary Task-based Methods}
In \cite{yang2021multi}, all unlabeled data generated by the losing bids are leveraged to estimate CTR to address the SSB issue. They proposed the Multi-task Advertising Estimator (MTAE), a multi-task learning framework for CTR prediction and market price modeling. MTAE takes advantage of the sufficient bid prices of the full-volume bid requests and incorporates the auxiliary task of estimating the winning probability into the model for unbiased learning.

\subsubsection{Evaluation Metrics}
\begin{itemize}
    \item \textbf{Area Under Curve (AUC)}: This is one of the most widely used metrics for utility estimation model evaluation. It reflects the ranking ability and is formulated as:
\begin{equation}
\label{eq:auc}
\begin{aligned}
AUC = \frac{1}{|S_{+}||S_{-}|} \sum_{x^+ \in S_+} \sum_{x^{-} \in S_{-}} I(\phi(x^+)>\phi(x^{-}), 
 \end{aligned}
\end{equation}
where $S_{+}$ and $S_{-}$ are the positive samples and negative samples, respectively. And $|\cdot|$ denotes the number of, $\phi(\cdot)$ and $I(\cdot)$ are the estimation function and the indicator function, respectively. 
    \item  \textbf{Group Area Under Curve (GAUC)}: This metric is computed as follows. Firstly, partition the test data samples into several groups based on the unique user ID. Secondly, calculate the AUC in each group. Then, average the weighted AUC. This process could be formulated as:
 \begin{equation}
\label{eq:gauc}
\begin{aligned}
GAUC = \frac{\sum_u w_u AUC_u}{\sum_u w_u},  
 \end{aligned}
\end{equation}
where $w_u$ is the weight for user $u$ and is set as 1 in utility estimation, and $AUC_u$ is user $u$'s AUC. 
    \item  \textbf{F1 Score (F1)}: It is formulated as:
 \begin{equation}
\label{eq:f1}
\begin{aligned}
F1 = \frac{TP}{TP+\frac{1}{2}(FP +FN)},  
 \end{aligned}
\end{equation}
where $FN$, $FP$, and $TP$ are the number of false negative, false positive and true positive estimations. 
    \item \textbf{Mean Squared Error (MSE)}: MSE is adoped in \cite{bao2020gmcm} to evaluate the effectiveness of their utility estimation function, which is defined in Eq. \eqref{eq:mse}.
    \item \textbf{Negative Log-likelihood}: Its formulation is defined as adding a negative symbol to that of Log-likelihood defined in Eq. \eqref{eq:ll}.
\end{itemize}

\subsubsection{Summary}
Sampling-based methods are faced with the numerical instability problem when introducing rejection sampling. For those based on sequential patterns of user response actions are concerned, despite performance improvement, the majority of them are still restricted to impression-level inference spaces. In addition, there is a lack of theoretical support about them being unbiased estimators. 
Similar to Sec. \ref{sec:wps_methods_summary}, auxiliary task-based methods face the problem of tuning hyperparameters, which is tedious. 

%To accurately predict the utility in the entire space, some factors or technologies could be taken into consideration:
%\begin{itemize}
%    \item Causal graphs: When causal graph is promising to address the bias problem, it can also provide opportunities to give explanation from the strong causal paths in the graph.
%    \item The delayed feedback: Many conversions can only be observed after a relatively long and random delay since clicks happened, resulting in many false negative labels during training. In this sense, the time delay factor need to be incorporated into the CVR prediction. 
%    \item Multi-task learning: CVR prediction is largely similar to CTR prediction and these two tasks share the same feature spaces of both the user and item. Representation sharing across these two tasks is helpful. In addition, CVR prediction could also take advantage of task like winning probability estimation. 
%\end{itemize}

\subsection{Robustness against Bid Shading Biases}
Before 2017, the second-price auction mechanism, in which the winner pays the second highest bid price to the supply side platforms, was the dominant type of AIRTB auction mechanisms. However, almost all the mainstream SSPs and ad exchange networks (e.g., OpenX, Rubicon Project, Pubmatic, Index Exchange, AppNexus) are starting to implement first-price auctions \cite{sarah2017, zhou2021efficient}. Two major reasons for moving away from second-price auction are: 1) as the bidders always pay exactly what they bid, first-price auction increases the transparency and accountability for the bidders \cite{chari1992us, getintent2017, rubicon2018, ssluis2017}; and 2) the widely used and favored techniques of Header Bidding are incompatible with the unaltered second-price auction mechanism \cite{bhovaness2018}.

DSPs can estimate their potential competitors' bid prices, and take their behaviors into consideration in order to precisely design the optimal bidding strategies in the first-price auction mechanism. If the DSP knows the competitors' bid prices in advance, providing the bid price slightly higher than the highest bid price from all competitors would be the optimal bidding strategy (i.e., winning the corresponding auction with the lowest bid price). However, this is impossible in practice, where the DSP needs to predict the minimum winning price precisely, and attempts to lower their original bid prices which they had set for the second-price auction mechanism, that means shading the true value of the inventory in accordance. Such a process is named as bid shading, which has been utilized in auctions across various industries but is new to online advertising, especially to AIRTB. 

In first-price auction, the minimum winning price is either the highest bid price submitted by competing advertisers or the floor price. However, whether or not to announce the minimum winning price after the auction is determined by the SSP. If the minimum winning price is announced, the first-price auction mechanism is referred to as open (or non-censored); otherwise, it is referred to as closed (or censored). 
If the auction is closed, DSPs need to estimate the winning prices. This is similar to winning probability estimation in second-price auction, and is also biased. 

%\subsection{Methods used to mitigate the bias in bid shading}

To the best of our knowledge, only one study \cite{zhou2021efficient} deals with the censored problem in first-price auction-based AIRTB. Specifically, they first utilize the formulation in \cite{pan2020bid} to formulate the cumulative distribution function of the minimum winning price, enhancing the flexibility to select distributions. Then, they obtain the parameters of the cumulative distribution function through minimizing the loss between the ground truth and the predicted likelihood of winning.

% \subsection{Nonbias Evaluation Metrics for winning probability estimation}
% There are 
% \begin{itemize}
%     \item \textbf{Area Under Curve (AUC)}: 
% \end{itemize}

\section{Fairness}
\label{sec:fairness}
Fairness, which is defined as ``the absence of any prejudice or favoritism towards an individual or a group based on their intrinsic or acquired traits in the context of decision making'' \cite{saxena2019fairness, mehrabi2021survey, liu2021trustworthy} is a crucial issue in AIRTB systems.
% Firstly,  from the standpoint of ethics, fairness is regarded as one of the essential attributes which make people’s lives better in ancient Greece \cite{ameriks2000aristotle}. Fairness is a crucial quality and a prerequisite for a just society \cite{rawls2004theory}. Secondly,  from the standpoint of legality, under the terms of Anti-discrimination laws \cite{holmes2005anti}, no one or group could be discriminated against in the workplace, in admissions, in housing, or in public services based on their gender, age, color, or other characteristics. Take it for example, in a job-opportunity ads delivery scenario, minority-owned businesses ought to receive the same number of impressions as white-owned businesse \cite{liu2020balancing}. In addition,
% From the standpoint of a user, a fair RTB ecosystem  promotes the inclusion of a variety of information, including some niche information, in the RTB ecosystem. This may break the information silo, lessen societal division, extend users' horizons and increase the revenue of the ecosystem. 
Ad creatives expect the AIRTB system offers fair exposure opportunities to them, avoiding the Matthew effect \cite{li2021user}. Being fair might also attract advertisers of niche ads or items, which can increase the variety and originality of the ads or products in an AIRTB system. 
From the perspective of the AIRTB ecosystem, fairness is advantageous over the long run. For instance, the unfair trading ecosystem might give more exposure opportunities to advertisers with short-lived popular items, leading to loss of users over time. Similarly, it might provide some niche providers with few impressions. Niche advertisers may have a propensity to leave unfair AIRTB systems as a result of the negative feedbacks, reducing the diversity of ads offered to users. 
Fairness can help improve users' loyalty to the AIRTB systems. In a nutshell, improving fairness is of vital importance for AIRTB systems.

Depending on the focus, fairness can be separated into two main categories: 1) process fairness and 2) outcome fairness \cite{wang2022survey}. Process fairness (a.k.a. procedural justice \cite{lee2019procedural}) requires fair allocation in the process \cite{lee2019procedural, narayanan2018translation}, while outcome fairness (a.k.a. distributive justice \cite{lee2019procedural}) requires fair outcomes as a result of fair allocation \cite{fang2020achieving, lee2019procedural}.
In the following part of this section, we pay more attention to outcome fairness as most of the state-of-the-art studies in AIRTB are anchored in this category. 
Outcome fairness can be achieved in two main ways: 1) grouped by the goal, and 2) grouped by the concept. Depending on the fairness level of the result, outcome fairness can be divided into individual fairness and group fairness. According to the fairness concept, outcome fairness includes various sub-categories, from those that have received much attention (e.g., consistent fairness, calibrated fairness) to those only explored by a small number of works (e.g., maximin-shared fairness, Rawlsian maximin fairness, envy-free fairness, counterfactual fairness). 

In order to ensure group fairness, two groups of individuals which have distinct sensitive attributes (e.g., gender, age, ethnicity) must statistically experience similar treatments and receive similar outcomes. 
% Having this in mind, a number of definitions have been put forth, including Equal Opportunity \cite{hardt2016equality}, Equal Odds \cite{hardt2016equality}, Demographic Parity \cite{dwork2012fairness}, the former of which stipulates that members of two groups must have an equal chance of succeeding when they do indeed fall into the category of success \cite{hardt2016equality}. Equal Odds stipulates that different groups must have an equal chance of being correctly classified \cite{hardt2016equality} while Demographic Parity stipulates that different groups must have an equal chance of succeeding \cite{dwork2012fairness}. 
% In addition, a group can be broken down into many subgroups when there are many fairness-related criteria. 
% Even though the groups under the attribute division of each particular attribute are fair, the subgroups may still treat one another unfairly. Therefore, fairness should be taken into account in these subgroups \cite{kearns2018preventing, kearns2019empirical}. 
Fair outcomes for a group of people can be ensured by group fairness. However, at the individual level, discrimination can still occur \cite{dwork2012fairness}. Individual fairness requires that outcomes must be equitable for each individual. In some contexts, individual fairness means that similar people should be treated equally \cite{dwork2012fairness, biega2018equity}. However, there are plenty of alternative ways to define individual-level fairness. In order to be clear, following work \cite{wang2022survey}, we refer to a broader meaning of fairness (i.e., the individual level fairness) as individual fairness.
It is worth noting that compared to individual fairness, group fairness is more complicated since there can be multiple divisions and these divisions may evolve over time, allowing one person to be a member of multiple groups at once \cite{ge2021towards}. In addition, individual fairness in scenarios in which each person belongs to a distinct group can be conceptually viewed as a specific example of group fairness. 

Unfair practices can ocurr in AIRTB systems. In the following, we first discuss unfairness in AIRTB. Then, we analyse the reasons behind the unfairness in AIRTB. Finally, we discuss existing studies proposed to mitigate unfairness in AIRTB.

\subsection{Unfairness in AIRTB Systems}
Unfairness can occur in AIRTB systems unintentionally and intentionally. For example, \cite{datta2014automated} shows that men were shown more high-paying job ads than women with comparable profiles. In addition, \cite{lambrecht2019algorithmic} experimentally demonstrated that STEM (science, technology, engineering and math) job ads, which are designed to be gender neutral, are displayed to fewer women than men across almost all major platforms. For instance, in Facebook, a platform where women make up 52\% \cite{vermeren2015} of the user population and who are more likely to click ads, women are far less likely to be shown such ads than men. The study found that women are a prized demographic, making them more expensive to advertise to. This implies that ads that are meant to be gender-neutral can be delivered in the way that appears to be discriminatory by AIRTB algorithms that focus on optimizing cost-effectiveness. 
Ali et al. \cite{ali2019discrimination} explained that this is not solely the indication of the ingrained cultural bias nor a result of user profiles inputted into ads algorithms, but rather the product of competitive spillovers among advertisers. 
%Due to such unfairness, Facebook has recently become the target of two cases that could result in civil lawsuits for alleged housing and job discrimination \cite{guynnj2018, timberg2018, nfha2018, angwin2016}. 
Apart from users, advertisers can also be unfairly treated during the AIRTB auction process \cite{turkel2022regulating}. 

Recently, several research works have devoted considerable effort to investigate the causes of unfairness in AIRTB. A number of possible explanations have been found.
On the one hand, the goal of AIRTB systems is to deliver the right ads to right users at the right time. To achieve such goal, the system creates detailed user profiles and monitors ads performance to learn how users respond to various ads. Based on historical data, subsequent ads can be delivered to targeted to users more precisely. However, during this process, the AIRTB system can unintentionally deliver ads primarily to certain groups of users. This is especially troubling when it comes to employment, housing, and credit related ads because unfairness in these categories can violate certain legislation.
On the other hand, market forces and financial optimization strongly influence how ads are delivered, as some user groups are bound to be more valuable than others \cite{dwork2018fairness, larrykim2011, liu2014measurement, saez2014beyond}. 
%As found in \cite{larrykim2011, liu2014measurement, saez2014beyond}, some users tend to be more valuable to advertisers than others. 
Therefore, advertisers on a tight budget are more likely to lose bids for the ``valued'' customers. In this case, if the sensitive attributes of these ``valued'' customers belong to the protected classes, this can result in unfair ad delivery, even if the advertisers may not intend to exclude these customers. 

\subsection{Enhancing Fairness in AIRTB Systems}
\begin{figure}[ht]
\centering
\includegraphics[width=1\columnwidth]{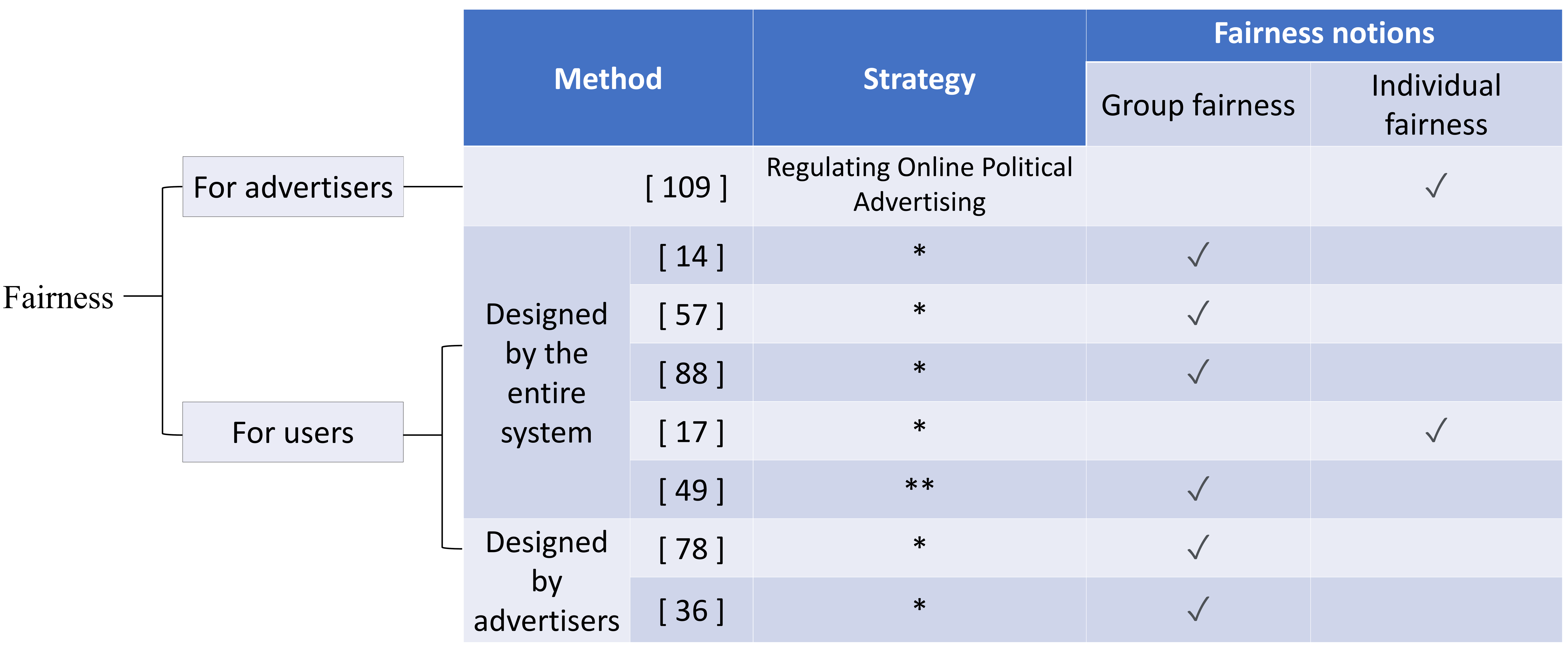}
\caption{Summary of methods used to ensure fairness in the RTB. * indicates that the method ensures fairness by formulating the fairness requirements as the constraints on the direct objective of maximizing the stakeholders' revenue. ** indicates that the method takes optimizing fairness as the direct objective.
}
\label{fig:fairness_summary}
\end{figure}
Fig. \ref{fig:fairness_summary} summarizes the methods used to mitigate unfairness in AIRTB systems.
Existing studies can be divided into two categories: 1) for advertisers, and 2) for users. The first category focuses on mitigating ad exchange networks' unfair favor for certain groups of advertisers. The second category focuses on fulfilling restrictions on the percentage of ad impressions that must reach particular groups of users belonging to certain demographics to enhance fair treatment of users. 

\subsubsection{Fairness for Advertisers} As there can be many advertisers connected to a specific ad exchange network, it is critical for the network to decide on which groups of advertisers to send bid requests to and how to choose winners for bid requests fairly in order to maximize its profit and build trust with advertisers. In \cite{turkel2022regulating}, it has been suggested that this goal can be achieved through basic modifications to the ad auction mechanism that the ad exchange network uses. However, this work focuses on providing theoretical models for regulating online ad auctions instead of mitigating unfairness. Enhancing fairness is just one of the aspects studied. As such, whether such intervention works in existing AIRTB systems still needs to be validated. 

\subsubsection{Fairness for Users}
Works falling into this category can be further divided into two groups: 1) methods designed by the entire system, and 2) methods designed by the advertisers. 

\textbf{Methods designed by the entire system.}
To prevent discriminatory advertisements with respect to sensitive attributes, \cite{celis2019toward} proposed an optimization-based ad auction framework based on Myerson auction \cite{myerson1981optimal}, which maximizes the revenue of the ecosystem conditioned on constraints that preclude the development of unintentional discrimination. The constraints can be any notion of group fairness described in \cite{celis2019controlling}. 
Works \cite{kuo2020proportionnet, peri2021preferencenet} extended \cite{celis2019toward} by adapting it into different auction frameworks and capturing requirements of fairness from data, respectively. 
% Towards getting the optimal parameters of the auction framework, which turns out to be a non-convex problem, the authors first transform it into a a mixture of a non-convex function which is unconstrained with a convex function that is confined to a polytope, and then adopt the gradient descent based scheme to solve these problems. 
Celis et al. \cite{celis2019toward} focuses on solving the optimization problem formulated based on algorithmic fairness, while \cite{chawla2022individual} focuses on comparing its performance with the unfair optimum and studying the cost of achieving fairness.
It is an individual fairness approach (i.e., any two individuals must obtain additively similar allocations from each advertiser if they have been given multiplicatively similar values by all the advertisers). They follow the idea of \cite{ilvento2020multi} and exam how auction design affects outcome fairness with the assumption that bids of each advertiser are accepted without discrimination. Afterwards, they adopt the Inverse Proportional Allocation algorithm to balance social welfare and fairness for a wide-range of value stability conditions. Apart from these methods which focus on algorithms design, \cite{imana2021auditing} proposed to adopt auditing to cope with group unfairness based on user features. However, this method depends greatly on human intervention and incurs high costs. 

\textbf{Methods designed by advertisers.}
Works under this category attempt to design bidding and targeting strategies that rectify the unfairness introduced by the AIRTB system from the perspective of the advertisers.
In \cite{nasr2020bidding}, the authors designed bidding strategies for advertisers to achieve impression parity across various demographic groups.
Similarly, in \cite{gelauff2020advertising}, the authors designed targeting strategies for achieving parity in outcomes or conversions across different user groups. In both approaches, the requirement of group fairness is formulated as the constraint on the main objective of maximizing advertisers' revenues. 

\subsection{Evaluation Metrics}
Statistical significance has been adopted to measure AIRTB fairness \cite{imana2021auditing}. Specifically, the percentages of people belong to different gender groups seeing two different ads are calculated. Then, the Z-test with a 95\% confidence level is performed to measure the statistical significance of the difference between these two groups.

%As far as the other existing methods used to guarantee fairness in the RTB auctioning ecosystem are concerned, as they formulate the fairness requirements as the constraints of maximizing the corresponding stakeholders revenue, they do not directly evaluate whether their methods truly ensure fairness among groups or individuals. 

\subsection{Summary}
It can be observed from Fig. \ref{fig:fairness_summary} that research on fair AIRTB techniques is still in its infancy. Existing works mostly focus on group fairness, while only two focusing on individual fairness \cite{chawla2022individual, turkel2022regulating}. Most of them formulate the fairness requirements as the constraints on the main objective function, with only one taking fairness as the goal \cite{imana2021auditing}. 

% \begin{figure}[t!]
% \centering
% \includegraphics[width=1\columnwidth]{image/roadmap.pdf}
% \caption{Roadmap of potential future works towards trustwothy AI-empowered RTB auctioning system.
% }
% \label{fig:roadmap_for_future_work}
% \end{figure}

\section{Promising Future Research Directions}
\label{sec:future_direction}
There are many areas where exciting new research can be carried out towards building a trustworthy AIRTB ecosystem. As shown in Table \ref{tab:requirements}, existing studies on trustworthy AIRTB focus on the security, robustness and fairness dimensions of trustworthy AI, leaving other dimensions such as accountability and explainability less well explored. In addition, there is still plenty of room for improvement for research on the security, robustness and fairness dimensions. In this section, we discuss promising future directions for this emerging interdisciplinary field.
%All these dimensions are crucial for building trust of various stakeholders towards the ecosystem. As such, more much efforts are needed to build such a trustworthy AIRTB system from perspectives of all stakeholders and all dimensions. 

\subsection{Security}
%\subsubsection{Combating attacks}
%According to the report from Juniper Research, more than \$42 billion has been lost by the online advertising industry due to attacks. By 2023, the sum is anticipated to reach \$100 billion. Therefore, combating attacks is still a critical task for the AIRTB system. According to this report, instead of employing strategies like app install farms, attackers frequently use techniques like domain spoofing to boost the click number by making websites with low quality appear to be those with high quality. Therefore, it is crucial to distinguish between real ad clicks and fake ones, which is difficult in the AIRTB system. 
% The core issues in guarding against various sorts of attacks to RTB auction outlined in Sec. \ref{sec:security} are the porous security of client-side softwares and the lack of transparency. As such, we will discuss some potential directions for the following works to cope with these problems. 
% In addition, as shown in Table \ref{tab:requirements}, the requirements of different major participants may be incompatible with one another Therefore, much works need to be done to simultaneously combat attacks against various stakeholders in the AIRTB system.  

\subsubsection{Demand for Transparency}
One problem of AIRTB systems is that it is difficult for the advertisers to acquire information about the sources of the traffics, leaving room for ad fraud and ad injection.
Typically, most participants may be hesitant to disclose the specifics of the approaches and techniques used for combating ad frauds. Since the advertisers are the sources of revenue and the foundation of the value chain for AIRTB systems, they have great incentive to help maintain the sustainability of the ecosystem. Therefore, techniques to help them more transparently assess the effectiveness of their ad campaigns and advertising strategies while guarding against ad frauds are desirable.

\subsubsection{Trade-offs between Stakeholders}
Table \ref{tab:requirements} lists the security requirements of various stakeholders of AIRTB. It can be observed that security requirements by different stakeholders may conflict. For instance, driven by profit, publishers may be tempted to generate fraudulent ad traffics, which contradicts with advertisers' requirements for a fraud free AIRTB system. 
% Moreover, to ensure the operation of the whole system, existing methods rely on collecting data from the users directly or indirectly. However, this may cause user data privacy and security concerns, violating the requirements of the users. 
To build the trustworthy AIRTB system, it is crucial to satisfy security the requirements by different stakeholders simultaneously. The research problem of striking the right balance among multiple AIRTB stakeholders' security requirements in a cost-effective manner remains open. 

\subsection{Robustness}
%In this section, we will discuss some open issues when mitigating bias in AIRTB and point out some potential directions for future works from the perspective of evaluation, framework, explanation and reasoning, etc. 
\subsubsection{Standardizing Evaluation Metrics}
To evaluate the effectiveness of methods enhancing the robustness of AIRTB systems, most existing works adopt MSE, ANLP and other conventional metrics which are designed to assess ranking performance. However, as shown in \cite{chen2020bias}, these metrics may not be well-suited for evaluating robustness. 
Furthermore, since different works perform evaluation following different metrics,
%more issues are present in the evaluation as a result of the popularity of bias. Generally, different existing works often use different evaluation metrics of unbias. 
the results are reported inconsistently. In order to objectively compare the performance of different approaches in this area of research, more suitable and standardized evaluation metrics need to be proposed.

%\subsubsection{Novel evaluation metrics}
%To build an unbiased AIRTB system, the state-of-the-art methods tend to depend on a large number of unbiased data, which degrades participants' experience due to concerns like privacy and is typically limited to a very tiny portion of online volume. To address this problem, some researchers tend to employ both small-scale unbiased data and large-scale biased data. However, such data are with high variance, making it difficult for existing adopted evaluation metrics to work well. As such, it will be fascinating to investigate novel evaluation metrics or evaluators to assist in the employment of both small-scale unbiased data and large-scale biased data. 

\subsubsection{Balancing Requirements from Various Stakeholders}
As shown in Table \ref{tab:requirements}, existing studies focus on satisfying the robustness requirements from the advertisers. As far as the publishers are concerned, in order for them to build trust with the AIRTB system, they also need to build unbiased models to optimally calculate the reserve price based on the historical auction records. It would be useful for future research to look into this area in order to serve the robustness needs of the publishers well. 

\subsubsection{Leveraging Auxiliary Information}
Taking advantage of the wealth of auxiliary information in AIRTB systems can enhance its robustness. There have been a few works in recent years showing that biases in recommender systems can be corrected by using user or item attributes. Since the ads, advertisers, publishers and users all possess auxiliary information, investigations on how to leverage such information to enhance the robustness of AIRTB systems can be worthwhile. 
\subsubsection{Reasoning}
Causal graphs among the stakeholders can be useful for enhancing the robustness of AIRTB systems. Reasoning about the occurrence and effect of bias are keys for debiasing. Causal graphs can provide a new source of information for this purpose. As such, building new reasoning techniques to leverage causal graphs in AIRTB to enhance its robustness against biases can be a promising future research direction.
%Causal graphs adopted by existing works in the field of AIRTB is simple and target at one specific case, making them inapplicable to various data and cases. As such, how to build better and more appropriate causal graphs and how to take advantage of them to help debiasing and reasoning may need more attention.

\subsection{Fairness}
%In this subsection, we present some possible directions for future  works towards the fair AIRTB systems from the perspectives of the definition, performance evaluation, and algorithm design.

%\subsubsection{Definition}
%There are various definitions of fairness, some of which have been discussed in this paper. Existing works in the AIRTB field only focus on group fairness and individual fairness. However, it is unclear what kind of results are fair. To cope with this issue, we may need more precise formal definitions of fairness and provide more information about fairness than the goal does. In this sense, it is crucial to introduce divergent views of fairness concepts, which describe what academics believe are the prerequisites for the fair results. 

\subsubsection{New Evaluation Metrics}
%1) \textbf{Fair comparison between various approaches}. 
%It's known that without reliable benchmarks, research in the corresponding area may suffer from unreliable assessment and unfair comparison, which negatively impact the growth of the scientific community. 
As shown in Section \ref{sec:fairness}, there is only one fairness metric used by existing studies in AIRTB. Compared to the diverse notions of fairness studied in this field, this is inadequate.
%Therefore, works need to evaluate their effectiveness by adopting the suitable metrics. In this process, as there are so many distinct fairness metrics and data-processing techniques, the state-of-the-art fairness studies suffer as a result. 
As such, new research on designing proper evaluation metrics to compare the fairness of AIRTB approaches is required.

\subsubsection{New Benchmarking Datasets} 
Existing methods tend to evaluate their methods using datasets collected by themselves from various advertising platforms, making it impossible to compare the performance in a standardized manner. There is a lack of public benchmarking datasets to study the fairness related approaches for AIRTB systems. This gap needs to be bridged in order for this field to be further advanced in a sustainable way.

%\subsubsection{Algorithm design}
%1) Most of existing works tend to formulate the fairness requirements, especially the group fairness requirement and individual fairness requirement, as the constraints of the goal of maximizing the revenue. However, such operation may make the requirements of fairness inflexible to adjust and change, and limit the incorporation of various fairness conceptions. As such, works specially designed for fairness are urgently needed. 

\subsubsection{Tradeoff between Fairness and Performance} 
In most cases, improving fairness means the loss in performance in machine learning. For example, in the field of recommender systems, the tradeoff between recommendation accuracy and fairness has been well studied \cite{wang2022survey}. However, fairness is not necessarily at odds with performance in well-designed systems. Particularly, in studies related to classification tasks \cite{lahoti2020fairness}, it is reported that increasing fairness may result in accuracy improvement. Therefore, investigating how to enhance fairness while maintaining performance for various stakeholders is crucial for successfully implementing fairness strategies in AIRTB systems.

\subsubsection{Multi-Faceted Fairness} 
Most of existing works focus on achieving just one notion of fairness and corresponding goals in AIRTB. However, as there are multiple fairness notions \cite{garcia2021maxmin, patro2020fairrec} and multiple stakeholders involved in AIRTB systems, it is useful to study how multi-faceted fairness can be achieved.

%4) \textbf{Fairness for the whole ecosystem}. Currently, there are various methods aiming at improving fairness in AIRTB systems, however the majority of them concentrate on the fairness from only one stakeholder's perspective, i.e., any of the advertisers and users but not both. However, fairness from all stakeholders' perspective is crucial for the development of the ecosystem and needs to be ensured. Hence, the proposals aimed at joint fairness may be worhtwhile. It is worth noting that there could be a disagreement between different stakeholders' fairness requirements, making the subject of joint fairness difficult. 

%5) \textbf{Fairness with other evaluation criteria}. Apart from accuracy and fairness, there exists various kinds of evaluation dimensions, including diversity, security, privacy and so on, to which different stakeholders' satisfaction is likewise strongly correlated. Consequently, in order to ensure user fairness in RTB as well as the fairness of other stakeholders, we should also take into account additional measurements further than accuracy.

\subsubsection{Causal Inference for Fairness} 
An emerging area of research in machine learning is on reducing unfairness through causal inference \cite{kusner2017counterfactual, wu2019counterfactual}. However, to our best knowledge, there is no study on fairness in AIRTB. Bridging this gap can lead to innovative new capabilities in the AIRTB system, enhancing its fairness in an interpretable manner.

\subsection{Accountability}
It is well-known that AIRTB systems have various drawbacks. Among them, the most noticeable one is that it can be challenging for the advertisers, publishers, DSPs and SSPs to assess their benefits derived from joining the auctions. For example, it is difficult for the advertisers to acquire information about the sources of ad traffics, leaving room for frauds and backroom deals. As such, it is necessary to design accountable and verifiable auction mechanisms to build trust with various stakeholders.

\section{Conclusions}
\label{sec:conclusion}
This paper provides a comprehensive review of the trustworthy AIRTB literature. To our best knowledge, this is the only survey on this emerging interdisciplinary area. We proposed a multi-tiered taxonomy to analyse trustworthy AIRTB works focusing on security, robustness and fairness. Under each topic, we summarize the challenges faced, the key approaches taken as well as the main evaluation metrics adopted to experimentally measure their performance in order to support long-term sustainability of this field. Finally, we suggest some promising future research directions that can help enhance trustworthiness of AIRTB systems. For this field to move forward, collaboration among researchers and industry practitioners is key. We hope that this survey can serve as a useful roadmap towards building trustworthy AIRTB systems of the future.

%Specifically, this paper reviews papers from three critical dimensions that contribute to trustworthy Real-Time Bidding advertising, security, robustness and fairness. For each dimension, we first make out the causality behind it and clarity the corresponding definitions; afterwards, we sort out studies used to achieve the goal of the dimension to further deepen the understanding of such dimension. Finally, we propose some potential research directions towards trusthworthy RTB auctioning ecosystem.

\section*{Acknowledgments}
This research is supported by the National Research Foundation, Singapore under its AI Singapore Programme (AISG Award No: AISG2-RP-2020-019); the Joint NTU-WeBank Research Centre on Fintech (Award No: NWJ-2020-008); the Nanyang Assistant Professorship (NAP); the RIE 2020 Advanced Manufacturing and Engineering (AME) Programmatic Fund (No. A20G8b0102), Singapore; and Future Communications Research \& Development Programme (FCP-NTU-RG-2021-014). Any opinions, findings and conclusions or recommendations expressed in this material are those of the author(s) and do not reflect the views of National Research Foundation, Singapore.

%%
%% The next two lines define the bibliography style to be used, and
%% the bibliography file.
%%% -*-BibTeX-*-
%%% Do NOT edit. File created by BibTeX with style
%%% ACM-Reference-Format-Journals [18-Jan-2012].

\bibliographystyle{ACM-Reference-Format}
\bibliography{main}

%%% -*-BibTeX-*-
%%% Do NOT edit. File created by BibTeX with style
%%% ACM-Reference-Format-Journals [18-Jan-2012].

\begin{thebibliography}{131}

%%% ====================================================================
%%% NOTE TO THE USER: you can override these defaults by providing
%%% customized versions of any of these macros before the \bibliography
%%% command.  Each of them MUST provide its own final punctuation,
%%% except for \shownote{}, \showDOI{}, and \showURL{}.  The latter two
%%% do not use final punctuation, in order to avoid confusing it with
%%% the Web address.
%%%
%%% To suppress output of a particular field, define its macro to expand
%%% to an empty string, or better, \unskip, like this:
%%%
%%% \newcommand{\showDOI}[1]{\unskip}   % LaTeX syntax
%%%
%%% \def \showDOI #1{\unskip}           % plain TeX syntax
%%%
%%% ====================================================================

\ifx \showCODEN    \undefined \def \showCODEN     #1{\unskip}     \fi
\ifx \showDOI      \undefined \def \showDOI       #1{#1}\fi
\ifx \showISBNx    \undefined \def \showISBNx     #1{\unskip}     \fi
\ifx \showISBNxiii \undefined \def \showISBNxiii  #1{\unskip}     \fi
\ifx \showISSN     \undefined \def \showISSN      #1{\unskip}     \fi
\ifx \showLCCN     \undefined \def \showLCCN      #1{\unskip}     \fi
\ifx \shownote     \undefined \def \shownote      #1{#1}          \fi
\ifx \showarticletitle \undefined \def \showarticletitle #1{#1}   \fi
\ifx \showURL      \undefined \def \showURL       {\relax}        \fi
% The following commands are used for tagged output and should be
% invisible to TeX
\providecommand\bibfield[2]{#2}
\providecommand\bibinfo[2]{#2}
\providecommand\natexlab[1]{#1}
\providecommand\showeprint[2][]{arXiv:#2}

\bibitem[li2(2012)]%
        {li2012knowing}
 \bibinfo{year}{2012}\natexlab{}.
\newblock \showarticletitle{Knowing your enemy: understanding and detecting
  malicious web advertising}. In \bibinfo{booktitle}{\emph{CSS}}.
  \bibinfo{pages}{674--686}.
\newblock


\bibitem[Aberathne and Walgampaya(2018)]%
        {aberathne2018smart}
\bibfield{author}{\bibinfo{person}{Iroshan Aberathne} {and}
  \bibinfo{person}{Chamila Walgampaya}.} \bibinfo{year}{2018}\natexlab{}.
\newblock \showarticletitle{Smart mobile bot detection through behavioral
  analysis}.
\newblock In \bibinfo{booktitle}{\emph{ICIIT}}. \bibinfo{pages}{241--252}.
\newblock


\bibitem[Ali et~al\mbox{.}(2019)]%
        {ali2019discrimination}
\bibfield{author}{\bibinfo{person}{Muhammad Ali}, \bibinfo{person}{Piotr
  Sapiezynski}, \bibinfo{person}{Miranda Bogen}, \bibinfo{person}{Aleksandra
  Korolova}, \bibinfo{person}{Alan Mislove}, {and} \bibinfo{person}{Aaron
  Rieke}.} \bibinfo{year}{2019}\natexlab{}.
\newblock \showarticletitle{Discrimination through optimization: How Facebook's
  Ad delivery can lead to biased outcomes}.
\newblock \bibinfo{journal}{\emph{Proceedings of the ACM on human-computer
  interaction}} \bibinfo{volume}{3}, \bibinfo{number}{CSCW}
  (\bibinfo{year}{2019}), \bibinfo{pages}{1--30}.
\newblock


\bibitem[Anupam et~al\mbox{.}(1999)]%
        {anupam1999security}
\bibfield{author}{\bibinfo{person}{Vinod Anupam}, \bibinfo{person}{Alain
  Mayer}, \bibinfo{person}{Kobbi Nissim}, \bibinfo{person}{Benny Pinkas}, {and}
  \bibinfo{person}{Michael~K Reiter}.} \bibinfo{year}{1999}\natexlab{}.
\newblock \showarticletitle{On the security of pay-per-click and other web
  advertising schemes}.
\newblock \bibinfo{journal}{\emph{Computer Networks}} \bibinfo{volume}{31},
  \bibinfo{number}{11-16} (\bibinfo{year}{1999}), \bibinfo{pages}{1091--1100}.
\newblock


\bibitem[B. Hovaness.(2018)]%
        {bhovaness2018}
B. Hovaness. \bibinfo{year}{2018}\natexlab{}.
\newblock \bibinfo{booktitle}{\emph{Sold for more than you should have paid}}.
\newblock
\urldef\tempurl%
\url{https://www.hearts-science.com/sold-for-more-than-you-should-have-paid/}
\showURL{%
Retrieved Jun 01, 2022 from \tempurl}


\bibitem[Badhe(2017)]%
        {badhe2017click}
\bibfield{author}{\bibinfo{person}{Anup Badhe}.}
  \bibinfo{year}{2017}\natexlab{}.
\newblock \showarticletitle{Click fraud detection in mobile ads served in
  programmatic inventory}.
\newblock \bibinfo{journal}{\emph{Neural Networks \& Machine Learning}}
  \bibinfo{volume}{1}, \bibinfo{number}{1} (\bibinfo{year}{2017}),
  \bibinfo{pages}{1--1}.
\newblock


\bibitem[Bao et~al\mbox{.}(2020)]%
        {bao2020gmcm}
\bibfield{author}{\bibinfo{person}{Wentian Bao}, \bibinfo{person}{Hong Wen},
  \bibinfo{person}{Sha Li}, \bibinfo{person}{Xiao-Yang Liu},
  \bibinfo{person}{Quan Lin}, {and} \bibinfo{person}{Keping Yang}.}
  \bibinfo{year}{2020}\natexlab{}.
\newblock \showarticletitle{Gmcm: Graph-based micro-behavior conversion model
  for post-click conversion rate estimation}. In
  \bibinfo{booktitle}{\emph{SIGIR}}. \bibinfo{pages}{2201--2210}.
\newblock


\bibitem[Berrar(2016)]%
        {berrar2016learning}
\bibfield{author}{\bibinfo{person}{Daniel Berrar}.}
  \bibinfo{year}{2016}\natexlab{}.
\newblock \showarticletitle{Learning from automatically labeled data: case
  study on click fraud prediction}.
\newblock \bibinfo{journal}{\emph{Knowledge and Information Systems}}
  \bibinfo{volume}{46}, \bibinfo{number}{2} (\bibinfo{year}{2016}),
  \bibinfo{pages}{477--490}.
\newblock


\bibitem[Biega et~al\mbox{.}(2018)]%
        {biega2018equity}
\bibfield{author}{\bibinfo{person}{Asia~J Biega}, \bibinfo{person}{Krishna~P
  Gummadi}, {and} \bibinfo{person}{Gerhard Weikum}.}
  \bibinfo{year}{2018}\natexlab{}.
\newblock \showarticletitle{Equity of attention: Amortizing individual fairness
  in rankings}. In \bibinfo{booktitle}{\emph{SIGIR}}.
  \bibinfo{pages}{405--414}.
\newblock


\bibitem[Brundage et~al\mbox{.}(2020)]%
        {brundage2020toward}
\bibfield{author}{\bibinfo{person}{Miles Brundage}, \bibinfo{person}{Shahar
  Avin}, \bibinfo{person}{Jasmine Wang}, \bibinfo{person}{Haydn Belfield},
  \bibinfo{person}{Gretchen Krueger}, \bibinfo{person}{Gillian Hadfield},
  \bibinfo{person}{Heidy Khlaaf}, \bibinfo{person}{Jingying Yang},
  \bibinfo{person}{Helen Toner}, \bibinfo{person}{Ruth Fong}, {et~al\mbox{.}}}
  \bibinfo{year}{2020}\natexlab{}.
\newblock \showarticletitle{Toward trustworthy AI development: mechanisms for
  supporting verifiable claims}.
\newblock \bibinfo{journal}{\emph{arXiv:2004.07213}} (\bibinfo{year}{2020}).
\newblock


\bibitem[Cai et~al\mbox{.}(2020)]%
        {cai2020threats}
\bibfield{author}{\bibinfo{person}{Yegui Cai}, \bibinfo{person}{George~OM Yee},
  \bibinfo{person}{Yuan~Xiang Gu}, {and} \bibinfo{person}{Chung-Horng Lung}.}
  \bibinfo{year}{2020}\natexlab{}.
\newblock \showarticletitle{Threats to online advertising and countermeasures:
  A technical survey}.
\newblock \bibinfo{journal}{\emph{Digital Threats: Research and Practice}}
  \bibinfo{volume}{1}, \bibinfo{number}{2} (\bibinfo{year}{2020}),
  \bibinfo{pages}{1--27}.
\newblock


\bibitem[Callejo et~al\mbox{.}(2016)]%
        {callejo2016independent}
\bibfield{author}{\bibinfo{person}{Patricia Callejo}, \bibinfo{person}{Ruben
  Cuevas}, \bibinfo{person}{Angel Cuevas}, {and} \bibinfo{person}{Mikko
  Kotila}.} \bibinfo{year}{2016}\natexlab{}.
\newblock \showarticletitle{Independent auditing of online display advertising
  campaigns}. In \bibinfo{booktitle}{\emph{HotNets}}.
  \bibinfo{pages}{120--126}.
\newblock


\bibitem[Cao et~al\mbox{.}(2020)]%
        {cao2020adsherlock}
\bibfield{author}{\bibinfo{person}{Chenhong Cao}, \bibinfo{person}{Yi Gao},
  \bibinfo{person}{Yang Luo}, \bibinfo{person}{Mingyuan Xia},
  \bibinfo{person}{Wei Dong}, \bibinfo{person}{Chun Chen}, {and}
  \bibinfo{person}{Xue Liu}.} \bibinfo{year}{2020}\natexlab{}.
\newblock \showarticletitle{AdSherlock: efficient and deployable click fraud
  detection for Mobile applications}.
\newblock \bibinfo{journal}{\emph{IEEE Transactions on Mobile Computing}}
  \bibinfo{volume}{20}, \bibinfo{number}{4} (\bibinfo{year}{2020}),
  \bibinfo{pages}{1285--1297}.
\newblock


\bibitem[Celis et~al\mbox{.}(2019b)]%
        {celis2019toward}
\bibfield{author}{\bibinfo{person}{Elisa Celis}, \bibinfo{person}{Anay
  Mehrotra}, {and} \bibinfo{person}{Nisheeth Vishnoi}.}
  \bibinfo{year}{2019}\natexlab{b}.
\newblock \showarticletitle{Toward controlling discrimination in online ad
  auctions}. In \bibinfo{booktitle}{\emph{ICML}}. \bibinfo{pages}{4456--4465}.
\newblock


\bibitem[Celis et~al\mbox{.}(2019a)]%
        {celis2019controlling}
\bibfield{author}{\bibinfo{person}{L~Elisa Celis}, \bibinfo{person}{Sayash
  Kapoor}, \bibinfo{person}{Farnood Salehi}, {and} \bibinfo{person}{Nisheeth
  Vishnoi}.} \bibinfo{year}{2019}\natexlab{a}.
\newblock \showarticletitle{Controlling polarization in personalization: An
  algorithmic framework}. In \bibinfo{booktitle}{\emph{FAT}}.
  \bibinfo{pages}{160--169}.
\newblock


\bibitem[Chari and Weber(1992)]%
        {chari1992us}
\bibfield{author}{\bibinfo{person}{V Chari} {and} \bibinfo{person}{Robert
  Weber}.} \bibinfo{year}{1992}\natexlab{}.
\newblock \showarticletitle{How the US Treasury should auction its debt}.
\newblock \bibinfo{journal}{\emph{Federal Reserve Bank of Minneapolis Quarterly
  Review}} \bibinfo{volume}{16}, \bibinfo{number}{4} (\bibinfo{year}{1992}).
\newblock


\bibitem[Chawla and Jagadeesan(2022)]%
        {chawla2022individual}
\bibfield{author}{\bibinfo{person}{Shuchi Chawla} {and} \bibinfo{person}{Meena
  Jagadeesan}.} \bibinfo{year}{2022}\natexlab{}.
\newblock \showarticletitle{Individual fairness in advertising auctions through
  inverse proportionality}. In \bibinfo{booktitle}{\emph{ITCS}}.
  \bibinfo{pages}{42:1--42:21}.
\newblock


\bibitem[Chen et~al\mbox{.}(2020)]%
        {chen2020bias}
\bibfield{author}{\bibinfo{person}{Jiawei Chen}, \bibinfo{person}{Hande Dong},
  \bibinfo{person}{Xiang Wang}, \bibinfo{person}{Fuli Feng},
  \bibinfo{person}{Meng Wang}, {and} \bibinfo{person}{Xiangnan He}.}
  \bibinfo{year}{2020}\natexlab{}.
\newblock \showarticletitle{Bias and debias in recommender system: A survey and
  future directions}.
\newblock \bibinfo{journal}{\emph{arXiv:2010.03240}} (\bibinfo{year}{2020}).
\newblock


\bibitem[Crussell et~al\mbox{.}(2014)]%
        {crussell2014madfraud}
\bibfield{author}{\bibinfo{person}{Jonathan Crussell}, \bibinfo{person}{Ryan
  Stevens}, {and} \bibinfo{person}{Hao Chen}.} \bibinfo{year}{2014}\natexlab{}.
\newblock \showarticletitle{Madfraud: Investigating ad fraud in android
  applications}. In \bibinfo{booktitle}{\emph{MobiSys}}.
  \bibinfo{pages}{123--134}.
\newblock


\bibitem[Datta et~al\mbox{.}(2014)]%
        {datta2014automated}
\bibfield{author}{\bibinfo{person}{Amit Datta}, \bibinfo{person}{Michael~Carl
  Tschantz}, {and} \bibinfo{person}{Anupam Datta}.}
  \bibinfo{year}{2014}\natexlab{}.
\newblock \showarticletitle{Automated experiments on ad privacy settings: A
  tale of opacity, choice, and discrimination}.
\newblock \bibinfo{journal}{\emph{arXiv:1408.6491}} (\bibinfo{year}{2014}).
\newblock


\bibitem[Dave et~al\mbox{.}(2012)]%
        {dave2012measuring}
\bibfield{author}{\bibinfo{person}{Vacha Dave}, \bibinfo{person}{Saikat Guha},
  {and} \bibinfo{person}{Yin Zhang}.} \bibinfo{year}{2012}\natexlab{}.
\newblock \showarticletitle{Measuring and fingerprinting click-spam in ad
  networks}. In \bibinfo{booktitle}{\emph{SIGCOMM}}. \bibinfo{pages}{175--186}.
\newblock


\bibitem[Dave et~al\mbox{.}(2013)]%
        {dave2013viceroi}
\bibfield{author}{\bibinfo{person}{Vacha Dave}, \bibinfo{person}{Saikat Guha},
  {and} \bibinfo{person}{Yin Zhang}.} \bibinfo{year}{2013}\natexlab{}.
\newblock \showarticletitle{Viceroi: Catching click-spam in search ad
  networks}. In \bibinfo{booktitle}{\emph{CCS}}. \bibinfo{pages}{765--776}.
\newblock


\bibitem[Ding(2010)]%
        {ding2010hybrid}
\bibfield{author}{\bibinfo{person}{Xuhua Ding}.}
  \bibinfo{year}{2010}\natexlab{}.
\newblock \showarticletitle{A hybrid method to detect deflation fraud in
  cost-per-action online advertising}. In \bibinfo{booktitle}{\emph{Proceedings
  of the 8th international conference on Applied Cryptography and Network
  Security (ACNS'10)}}. \bibinfo{pages}{545--562}.
\newblock


\bibitem[Dong et~al\mbox{.}(2011)]%
        {dong2011adsentry}
\bibfield{author}{\bibinfo{person}{Xinshu Dong}, \bibinfo{person}{Minh Tran},
  \bibinfo{person}{Zhenkai Liang}, {and} \bibinfo{person}{Xuxian Jiang}.}
  \bibinfo{year}{2011}\natexlab{}.
\newblock \showarticletitle{AdSentry: comprehensive and flexible confinement of
  JavaScript-based advertisements}. In \bibinfo{booktitle}{\emph{ACSAC}}.
  \bibinfo{pages}{297--306}.
\newblock


\bibitem[Dritsoula and Musacchio(2014)]%
        {dritsoula2014game}
\bibfield{author}{\bibinfo{person}{Lemonia Dritsoula} {and}
  \bibinfo{person}{John Musacchio}.} \bibinfo{year}{2014}\natexlab{}.
\newblock \showarticletitle{A game of clicks: Economic incentives to fight
  click fraud in ad networks}.
\newblock \bibinfo{journal}{\emph{ACM SIGMETRICS Performance Evaluation
  Review}} \bibinfo{volume}{41}, \bibinfo{number}{4} (\bibinfo{year}{2014}),
  \bibinfo{pages}{12--15}.
\newblock


\bibitem[Dwork et~al\mbox{.}(2012)]%
        {dwork2012fairness}
\bibfield{author}{\bibinfo{person}{Cynthia Dwork}, \bibinfo{person}{Moritz
  Hardt}, \bibinfo{person}{Toniann Pitassi}, \bibinfo{person}{Omer Reingold},
  {and} \bibinfo{person}{Richard Zemel}.} \bibinfo{year}{2012}\natexlab{}.
\newblock \showarticletitle{Fairness through awareness}. In
  \bibinfo{booktitle}{\emph{ITCS}}. \bibinfo{pages}{214--226}.
\newblock


\bibitem[Dwork and Ilvento(2018)]%
        {dwork2018fairness}
\bibfield{author}{\bibinfo{person}{Cynthia Dwork} {and}
  \bibinfo{person}{Christina Ilvento}.} \bibinfo{year}{2018}\natexlab{}.
\newblock \showarticletitle{Fairness under composition}.
\newblock \bibinfo{journal}{\emph{arXiv:1806.06122}} (\bibinfo{year}{2018}).
\newblock


\bibitem[Estrada-Jim{\'e}nez et~al\mbox{.}(2017)]%
        {estrada2017online}
\bibfield{author}{\bibinfo{person}{Jos{\'e} Estrada-Jim{\'e}nez},
  \bibinfo{person}{Javier Parra-Arnau}, \bibinfo{person}{Ana
  Rodr{\'\i}guez-Hoyos}, {and} \bibinfo{person}{Jordi Forn{\'e}}.}
  \bibinfo{year}{2017}\natexlab{}.
\newblock \showarticletitle{Online advertising: Analysis of privacy threats and
  protection approaches}.
\newblock \bibinfo{journal}{\emph{Computer Communications}}
  \bibinfo{volume}{100}, \bibinfo{number}{1} (\bibinfo{year}{2017}),
  \bibinfo{pages}{32--51}.
\newblock


\bibitem[Fang et~al\mbox{.}(2020)]%
        {fang2020achieving}
\bibfield{author}{\bibinfo{person}{Boli Fang}, \bibinfo{person}{Miao Jiang},
  \bibinfo{person}{Pei-yi Cheng}, \bibinfo{person}{Jerry Shen}, {and}
  \bibinfo{person}{Yi Fang}.} \bibinfo{year}{2020}\natexlab{}.
\newblock \showarticletitle{Achieving Outcome Fairness in Machine Learning
  Models for Social Decision Problems.}. In \bibinfo{booktitle}{\emph{IJCAI}}.
  \bibinfo{pages}{444--450}.
\newblock


\bibitem[Felt et~al\mbox{.}(2011)]%
        {felt2011survey}
\bibfield{author}{\bibinfo{person}{Adrienne~Porter Felt},
  \bibinfo{person}{Matthew Finifter}, \bibinfo{person}{Erika Chin},
  \bibinfo{person}{Steve Hanna}, {and} \bibinfo{person}{David Wagner}.}
  \bibinfo{year}{2011}\natexlab{}.
\newblock \showarticletitle{A survey of mobile malware in the wild}. In
  \bibinfo{booktitle}{\emph{SPSM}}. \bibinfo{pages}{3--14}.
\newblock


\bibitem[Fleming and Harrington(2011)]%
        {fleming2011counting}
\bibfield{author}{\bibinfo{person}{Thomas~R Fleming} {and}
  \bibinfo{person}{David~P Harrington}.} \bibinfo{year}{2011}\natexlab{}.
\newblock \bibinfo{booktitle}{\emph{Counting processes and survival analysis}}.
\newblock \bibinfo{publisher}{John Wiley \& Sons}.
\newblock


\bibitem[Ford et~al\mbox{.}(2009)]%
        {ford2009analyzing}
\bibfield{author}{\bibinfo{person}{Sean Ford}, \bibinfo{person}{Marco Cova},
  \bibinfo{person}{Christopher Kruegel}, {and} \bibinfo{person}{Giovanni
  Vigna}.} \bibinfo{year}{2009}\natexlab{}.
\newblock \showarticletitle{Analyzing and detecting malicious flash
  advertisements}. In \bibinfo{booktitle}{\emph{ACSAC}}.
  \bibinfo{pages}{363--372}.
\newblock


\bibitem[Gandhi et~al\mbox{.}(2006)]%
        {gandhi2006badvertisements}
\bibfield{author}{\bibinfo{person}{Mona Gandhi}, \bibinfo{person}{Markus
  Jakobsson}, {and} \bibinfo{person}{Jacob Ratkiewicz}.}
  \bibinfo{year}{2006}\natexlab{}.
\newblock \showarticletitle{Badvertisements: Stealthy click-fraud with
  unwitting accessories}.
\newblock \bibinfo{journal}{\emph{Journal of Digital Forensic Practice}}
  \bibinfo{volume}{1}, \bibinfo{number}{2} (\bibinfo{year}{2006}),
  \bibinfo{pages}{131--142}.
\newblock


\bibitem[Garc{\'\i}a-Soriano and Bonchi(2021)]%
        {garcia2021maxmin}
\bibfield{author}{\bibinfo{person}{David Garc{\'\i}a-Soriano} {and}
  \bibinfo{person}{Francesco Bonchi}.} \bibinfo{year}{2021}\natexlab{}.
\newblock \showarticletitle{Maxmin-fair ranking: individual fairness under
  group-fairness constraints}. In \bibinfo{booktitle}{\emph{KDD}}.
  \bibinfo{pages}{436--446}.
\newblock


\bibitem[Ge et~al\mbox{.}(2021)]%
        {ge2021towards}
\bibfield{author}{\bibinfo{person}{Yingqiang Ge}, \bibinfo{person}{Shuchang
  Liu}, \bibinfo{person}{Ruoyuan Gao}, \bibinfo{person}{Yikun Xian},
  \bibinfo{person}{Yunqi Li}, \bibinfo{person}{Xiangyu Zhao},
  \bibinfo{person}{Changhua Pei}, \bibinfo{person}{Fei Sun},
  \bibinfo{person}{Junfeng Ge}, \bibinfo{person}{Wenwu Ou}, {et~al\mbox{.}}}
  \bibinfo{year}{2021}\natexlab{}.
\newblock \showarticletitle{Towards long-term fairness in recommendation}. In
  \bibinfo{booktitle}{\emph{WSDM}}. \bibinfo{pages}{445--453}.
\newblock


\bibitem[Gelauff et~al\mbox{.}(2020)]%
        {gelauff2020advertising}
\bibfield{author}{\bibinfo{person}{Lodewijk Gelauff}, \bibinfo{person}{Ashish
  Goel}, \bibinfo{person}{Kamesh Munagala}, {and} \bibinfo{person}{Sravya
  Yandamuri}.} \bibinfo{year}{2020}\natexlab{}.
\newblock \showarticletitle{Advertising for demographically fair outcomes}.
\newblock \bibinfo{journal}{\emph{arXiv:2006.03983}} (\bibinfo{year}{2020}).
\newblock


\bibitem[Getintent.(2017)]%
        {getintent2017}
Getintent. \bibinfo{year}{2017}\natexlab{}.
\newblock \bibinfo{booktitle}{\emph{RTB Auctions: Fair Play?}}
\newblock
\urldef\tempurl%
\url{https://blog.getintent.com/rtb-auctions-fair-play-3b372d505089}
\showURL{%
Retrieved Jun 01, 2022 from \tempurl}


\bibitem[Ghosh et~al\mbox{.}(2019)]%
        {ghosh2019scalable}
\bibfield{author}{\bibinfo{person}{Aritra Ghosh}, \bibinfo{person}{Saayan
  Mitra}, \bibinfo{person}{Somdeb Sarkhel}, \bibinfo{person}{Jason Xie},
  \bibinfo{person}{Gang Wu}, {and} \bibinfo{person}{Viswanathan Swaminathan}.}
  \bibinfo{year}{2019}\natexlab{}.
\newblock \showarticletitle{Scalable bid landscape forecasting in real-time
  bidding}. In \bibinfo{booktitle}{\emph{ECML PKDD}}.
  \bibinfo{pages}{451--466}.
\newblock


\bibitem[Goel et~al\mbox{.}(2010)]%
        {goel2010understanding}
\bibfield{author}{\bibinfo{person}{Manish~Kumar Goel}, \bibinfo{person}{Pardeep
  Khanna}, {and} \bibinfo{person}{Jugal Kishore}.}
  \bibinfo{year}{2010}\natexlab{}.
\newblock \showarticletitle{Understanding survival analysis: Kaplan-Meier
  estimate}.
\newblock \bibinfo{journal}{\emph{International journal of Ayurveda research}}
  \bibinfo{volume}{1}, \bibinfo{number}{4} (\bibinfo{year}{2010}),
  \bibinfo{pages}{274}.
\newblock


\bibitem[Grace et~al\mbox{.}(2012)]%
        {grace2012unsafe}
\bibfield{author}{\bibinfo{person}{Michael~C Grace}, \bibinfo{person}{Wu Zhou},
  \bibinfo{person}{Xuxian Jiang}, {and} \bibinfo{person}{Ahmad-Reza Sadeghi}.}
  \bibinfo{year}{2012}\natexlab{}.
\newblock \showarticletitle{Unsafe exposure analysis of mobile in-app
  advertisements}. In \bibinfo{booktitle}{\emph{WISEC}}.
  \bibinfo{pages}{101--112}.
\newblock


\bibitem[Greene(2005)]%
        {greene2005censored}
\bibfield{author}{\bibinfo{person}{William~H Greene}.}
  \bibinfo{year}{2005}\natexlab{}.
\newblock \showarticletitle{Censored data and truncated distributions}.
\newblock \bibinfo{journal}{\emph{Available at SSRN 825845}}
  (\bibinfo{year}{2005}).
\newblock


\bibitem[Haddadi(2010)]%
        {haddadi2010fighting}
\bibfield{author}{\bibinfo{person}{Hamed Haddadi}.}
  \bibinfo{year}{2010}\natexlab{}.
\newblock \showarticletitle{Fighting online click-fraud using bluff ads}.
\newblock \bibinfo{journal}{\emph{ACM SIGCOMM Computer Communication Review}}
  \bibinfo{volume}{40}, \bibinfo{number}{2} (\bibinfo{year}{2010}),
  \bibinfo{pages}{21--25}.
\newblock


\bibitem[Hildebrandt(2019)]%
        {hildebrandt2019privacy}
\bibfield{author}{\bibinfo{person}{Mireille Hildebrandt}.}
  \bibinfo{year}{2019}\natexlab{}.
\newblock \showarticletitle{Privacy as protection of the incomputable self:
  From agnostic to agonistic machine learning}.
\newblock \bibinfo{journal}{\emph{Theoretical Inquiries in Law}}
  \bibinfo{volume}{20}, \bibinfo{number}{1} (\bibinfo{year}{2019}),
  \bibinfo{pages}{83--121}.
\newblock


\bibitem[Huang et~al\mbox{.}(2017)]%
        {huang2017game}
\bibfield{author}{\bibinfo{person}{Chin-Tser Huang},
  \bibinfo{person}{Muhammad~N Sakib}, \bibinfo{person}{Charles Kamhoua},
  \bibinfo{person}{Kevin Kwiat}, {and} \bibinfo{person}{Laurent Njilla}.}
  \bibinfo{year}{2017}\natexlab{}.
\newblock \showarticletitle{A game theoretic approach for inspecting web-based
  malvertising}. In \bibinfo{booktitle}{\emph{ICC}}. \bibinfo{pages}{1--6}.
\newblock


\bibitem[Huang et~al\mbox{.}(2018)]%
        {huang2018bayesian}
\bibfield{author}{\bibinfo{person}{Chin-Tser Huang},
  \bibinfo{person}{Muhammad~N Sakib}, \bibinfo{person}{Charles~A Kamhoua},
  \bibinfo{person}{Kevin~A Kwiat}, {and} \bibinfo{person}{Laurent Njilla}.}
  \bibinfo{year}{2018}\natexlab{}.
\newblock \showarticletitle{A bayesian game theoretic approach for inspecting
  web-based malvertising}.
\newblock \bibinfo{journal}{\emph{IEEE Transactions on Dependable and Secure
  Computing}} \bibinfo{volume}{17}, \bibinfo{number}{6} (\bibinfo{year}{2018}),
  \bibinfo{pages}{1257--1268}.
\newblock


\bibitem[IAB.(2016)]%
        {iab2016}
IAB. \bibinfo{year}{2016}\natexlab{}.
\newblock \bibinfo{booktitle}{\emph{IAB Tech Lab Publisher Ad Blocking
  Primer}}.
\newblock
\urldef\tempurl%
\url{https://www.iab.com/wp-content/uploads/2016/03/IABTechLab_Publisher_AdBlocking_Primer.pdf}
\showURL{%
Retrieved Jun 01, 2022 from \tempurl}


\bibitem[Ilvento et~al\mbox{.}(2020)]%
        {ilvento2020multi}
\bibfield{author}{\bibinfo{person}{Christina Ilvento}, \bibinfo{person}{Meena
  Jagadeesan}, {and} \bibinfo{person}{Shuchi Chawla}.}
  \bibinfo{year}{2020}\natexlab{}.
\newblock \showarticletitle{Multi-category fairness in sponsored search
  auctions}. In \bibinfo{booktitle}{\emph{FAT}}. \bibinfo{pages}{348--358}.
\newblock


\bibitem[Imana et~al\mbox{.}(2021)]%
        {imana2021auditing}
\bibfield{author}{\bibinfo{person}{Basileal Imana}, \bibinfo{person}{Aleksandra
  Korolova}, {and} \bibinfo{person}{John Heidemann}.}
  \bibinfo{year}{2021}\natexlab{}.
\newblock \showarticletitle{Auditing for discrimination in algorithms
  delivering job ads}. In \bibinfo{booktitle}{\emph{WWW}}.
  \bibinfo{pages}{3767--3778}.
\newblock


\bibitem[Iqbal et~al\mbox{.}(2018b)]%
        {iqbal2018protecting}
\bibfield{author}{\bibinfo{person}{Md~Shahrear Iqbal},
  \bibinfo{person}{Mohammad Zulkernine}, \bibinfo{person}{Fehmi Jaafar}, {and}
  \bibinfo{person}{Yuan Gu}.} \bibinfo{year}{2018}\natexlab{b}.
\newblock \showarticletitle{Protecting Internet users from becoming victimized
  attackers of click-fraud}.
\newblock \bibinfo{journal}{\emph{Journal of Software: Evolution and Process}}
  \bibinfo{volume}{30}, \bibinfo{number}{3} (\bibinfo{year}{2018}),
  \bibinfo{pages}{e1871}.
\newblock


\bibitem[Iqbal et~al\mbox{.}(2018a)]%
        {iqbal2018adgraph}
\bibfield{author}{\bibinfo{person}{Umar Iqbal}, \bibinfo{person}{Zubair
  Shafiq}, \bibinfo{person}{Peter Snyder}, \bibinfo{person}{Shitong Zhu},
  \bibinfo{person}{Zhiyun Qian}, {and} \bibinfo{person}{Benjamin Livshits}.}
  \bibinfo{year}{2018}\natexlab{a}.
\newblock \showarticletitle{Adgraph: A machine learning approach to automatic
  and effective adblocking}.
\newblock \bibinfo{journal}{\emph{arXiv:1805.09155}} (\bibinfo{year}{2018}).
\newblock


\bibitem[J. Chen, D. Lin, A. Kaufman, and Y. Villa.(2014)]%
        {chenj2014}
J. Chen, D. Lin, A. Kaufman, and Y. Villa. \bibinfo{year}{2014}\natexlab{}.
\newblock \bibinfo{booktitle}{\emph{Click stream analysis for fraud
  detection}}.
\newblock
\urldef\tempurl%
\url{https://patents.google.com/patent/US8880441}
\showURL{%
Retrieved Jun 01, 2022 from \tempurl}


\bibitem[Juels et~al\mbox{.}(2007)]%
        {juels2007combating}
\bibfield{author}{\bibinfo{person}{Ari Juels}, \bibinfo{person}{Sid Stamm},
  {and} \bibinfo{person}{Markus Jakobsson}.} \bibinfo{year}{2007}\natexlab{}.
\newblock \showarticletitle{Combating Click Fraud via Premium Clicks.}. In
  \bibinfo{booktitle}{\emph{USENIX Security}}. \bibinfo{pages}{17--26}.
\newblock


\bibitem[Kamar et~al\mbox{.}(2012)]%
        {kamar2012combining}
\bibfield{author}{\bibinfo{person}{Ece Kamar}, \bibinfo{person}{Severin
  Hacker}, {and} \bibinfo{person}{Eric Horvitz}.}
  \bibinfo{year}{2012}\natexlab{}.
\newblock \showarticletitle{Combining human and machine intelligence in
  large-scale crowdsourcing.}. In \bibinfo{booktitle}{\emph{AAMAS}},
  Vol.~\bibinfo{volume}{12}. \bibinfo{pages}{467--474}.
\newblock


\bibitem[Kaplan and Meier(1958)]%
        {kaplan1958nonparametric}
\bibfield{author}{\bibinfo{person}{Edward~L Kaplan} {and} \bibinfo{person}{Paul
  Meier}.} \bibinfo{year}{1958}\natexlab{}.
\newblock \showarticletitle{Nonparametric estimation from incomplete
  observations}.
\newblock \bibinfo{journal}{\emph{Journal of the American statistical
  association}} \bibinfo{volume}{53}, \bibinfo{number}{282}
  (\bibinfo{year}{1958}), \bibinfo{pages}{457--481}.
\newblock


\bibitem[Kshetri and Voas(2019)]%
        {kshetri2019online}
\bibfield{author}{\bibinfo{person}{Nir Kshetri} {and} \bibinfo{person}{Jeffrey
  Voas}.} \bibinfo{year}{2019}\natexlab{}.
\newblock \showarticletitle{Online advertising fraud}.
\newblock \bibinfo{journal}{\emph{Computer}} \bibinfo{volume}{52},
  \bibinfo{number}{1} (\bibinfo{year}{2019}), \bibinfo{pages}{58--61}.
\newblock


\bibitem[Kuo et~al\mbox{.}(2020)]%
        {kuo2020proportionnet}
\bibfield{author}{\bibinfo{person}{Kevin Kuo}, \bibinfo{person}{Anthony
  Ostuni}, \bibinfo{person}{Elizabeth Horishny}, \bibinfo{person}{Michael~J
  Curry}, \bibinfo{person}{Samuel Dooley}, \bibinfo{person}{Ping-yeh Chiang},
  \bibinfo{person}{Tom Goldstein}, {and} \bibinfo{person}{John~P Dickerson}.}
  \bibinfo{year}{2020}\natexlab{}.
\newblock \showarticletitle{Proportionnet: Balancing fairness and revenue for
  auction design with deep learning}.
\newblock \bibinfo{journal}{\emph{arXiv:2010.06398}} (\bibinfo{year}{2020}).
\newblock


\bibitem[Kusner et~al\mbox{.}({[n.\,d.]})]%
        {kusner2017counterfactual}
\bibfield{author}{\bibinfo{person}{Matt~J Kusner}, \bibinfo{person}{Joshua
  Loftus}, \bibinfo{person}{Chris Russell}, {and} \bibinfo{person}{Ricardo
  Silva}.} \bibinfo{year}{[n.\,d.]}\natexlab{}.
\newblock \showarticletitle{Counterfactual fairness}.
\newblock \bibinfo{journal}{\emph{NIPS}}, \bibinfo{pages}{4066--4076}.
\newblock


\bibitem[Lahoti et~al\mbox{.}(2020)]%
        {lahoti2020fairness}
\bibfield{author}{\bibinfo{person}{Preethi Lahoti}, \bibinfo{person}{Alex
  Beutel}, \bibinfo{person}{Jilin Chen}, \bibinfo{person}{Kang Lee},
  \bibinfo{person}{Flavien Prost}, \bibinfo{person}{Nithum Thain},
  \bibinfo{person}{Xuezhi Wang}, {and} \bibinfo{person}{Ed Chi}.}
  \bibinfo{year}{2020}\natexlab{}.
\newblock \showarticletitle{Fairness without demographics through adversarially
  reweighted learning}.
\newblock \bibinfo{journal}{\emph{NIPS}}, \bibinfo{pages}{728--740}.
\newblock


\bibitem[Lambrecht and Tucker(2019)]%
        {lambrecht2019algorithmic}
\bibfield{author}{\bibinfo{person}{Anja Lambrecht} {and}
  \bibinfo{person}{Catherine Tucker}.} \bibinfo{year}{2019}\natexlab{}.
\newblock \showarticletitle{Algorithmic bias? An empirical study of apparent
  gender-based discrimination in the display of STEM career ads}.
\newblock \bibinfo{journal}{\emph{Management science}} \bibinfo{volume}{65},
  \bibinfo{number}{7} (\bibinfo{year}{2019}), \bibinfo{pages}{2966--2981}.
\newblock


\bibitem[Larry Kim.(2011)]%
        {larrykim2011}
Larry Kim. \bibinfo{year}{2011}\natexlab{}.
\newblock \bibinfo{booktitle}{\emph{The Most Expensive Keywords In Google
  Adwords}}.
\newblock
\urldef\tempurl%
\url{https://www.wordstream.com/blog/ws/2011/07/18/most-expensive-keywords-google-adwords}
\showURL{%
Retrieved Jun 01, 2022 from \tempurl}


\bibitem[Lee et~al\mbox{.}(2019)]%
        {lee2019procedural}
\bibfield{author}{\bibinfo{person}{Min~Kyung Lee}, \bibinfo{person}{Anuraag
  Jain}, \bibinfo{person}{Hea~Jin Cha}, \bibinfo{person}{Shashank Ojha}, {and}
  \bibinfo{person}{Daniel Kusbit}.} \bibinfo{year}{2019}\natexlab{}.
\newblock \showarticletitle{Procedural justice in algorithmic fairness:
  Leveraging transparency and outcome control for fair algorithmic mediation}.
\newblock \bibinfo{journal}{\emph{Proceedings of the ACM on Human-Computer
  Interaction}} \bibinfo{volume}{3}, \bibinfo{number}{CSCW}
  (\bibinfo{year}{2019}), \bibinfo{pages}{1--26}.
\newblock


\bibitem[Li et~al\mbox{.}(2015)]%
        {li2015adattester}
\bibfield{author}{\bibinfo{person}{Wenhao Li}, \bibinfo{person}{Haibo Li},
  \bibinfo{person}{Haibo Chen}, {and} \bibinfo{person}{Yubin Xia}.}
  \bibinfo{year}{2015}\natexlab{}.
\newblock \showarticletitle{Adattester: Secure online mobile advertisement
  attestation using trustzone}. In \bibinfo{booktitle}{\emph{MobiSys}}.
  \bibinfo{pages}{75--88}.
\newblock


\bibitem[Li et~al\mbox{.}(2014)]%
        {li2014search}
\bibfield{author}{\bibinfo{person}{Xin Li}, \bibinfo{person}{Min Zhang},
  \bibinfo{person}{Yiqun Liu}, \bibinfo{person}{Shaoping Ma},
  \bibinfo{person}{Yijiang Jin}, {and} \bibinfo{person}{Liyun Ru}.}
  \bibinfo{year}{2014}\natexlab{}.
\newblock \showarticletitle{Search engine click spam detection based on
  bipartite graph propagation}. In \bibinfo{booktitle}{\emph{WSDM}}.
  \bibinfo{pages}{93--102}.
\newblock


\bibitem[Li et~al\mbox{.}(2021)]%
        {li2021user}
\bibfield{author}{\bibinfo{person}{Yunqi Li}, \bibinfo{person}{Hanxiong Chen},
  \bibinfo{person}{Zuohui Fu}, \bibinfo{person}{Yingqiang Ge}, {and}
  \bibinfo{person}{Yongfeng Zhang}.} \bibinfo{year}{2021}\natexlab{}.
\newblock \showarticletitle{User-oriented fairness in recommendation}. In
  \bibinfo{booktitle}{\emph{WWW}}. \bibinfo{pages}{624--632}.
\newblock


\bibitem[Little and Rubin(2019)]%
        {little2019statistical}
\bibfield{author}{\bibinfo{person}{Roderick~JA Little} {and}
  \bibinfo{person}{Donald~B Rubin}.} \bibinfo{year}{2019}\natexlab{}.
\newblock \bibinfo{booktitle}{\emph{Statistical analysis with missing data}}.
  Vol.~\bibinfo{volume}{793}.
\newblock \bibinfo{publisher}{John Wiley \& Sons}.
\newblock


\bibitem[Liu et~al\mbox{.}(2014b)]%
        {liu2014decaf}
\bibfield{author}{\bibinfo{person}{Bin Liu}, \bibinfo{person}{Suman Nath},
  \bibinfo{person}{Ramesh Govindan}, {and} \bibinfo{person}{Jie Liu}.}
  \bibinfo{year}{2014}\natexlab{b}.
\newblock \showarticletitle{DECAF: Detecting and characterizing ad fraud in
  mobile apps}. In \bibinfo{booktitle}{\emph{NSDI}}. \bibinfo{pages}{57--70}.
\newblock


\bibitem[Liu et~al\mbox{.}(2022)]%
        {liu2021trustworthy}
\bibfield{author}{\bibinfo{person}{Haochen Liu}, \bibinfo{person}{Yiqi Wang},
  \bibinfo{person}{Wenqi Fan}, \bibinfo{person}{Xiaorui Liu},
  \bibinfo{person}{Yaxin Li}, \bibinfo{person}{Shaili Jain},
  \bibinfo{person}{Yunhao Liu}, \bibinfo{person}{Anil~K Jain}, {and}
  \bibinfo{person}{Jiliang Tang}.} \bibinfo{year}{2022}\natexlab{}.
\newblock \showarticletitle{Trustworthy ai: A computational perspective}.
\newblock \bibinfo{journal}{\emph{ACM Transactions on Intelligent Systems and
  Technology}} (\bibinfo{year}{2022}).
\newblock
\urldef\tempurl%
\url{https://doi.org/10.1145/3546872}
\showDOI{\tempurl}


\bibitem[Liu et~al\mbox{.}(2020)]%
        {liu2020research}
\bibfield{author}{\bibinfo{person}{Mengjuan Liu}, \bibinfo{person}{Wei Yue},
  \bibinfo{person}{Lizhou Qiu}, \bibinfo{person}{Jiaxing Li}, {and}
  \bibinfo{person}{Zhiguang Qin}.} \bibinfo{year}{2020}\natexlab{}.
\newblock \showarticletitle{Research Progress of Real-Time Bidding for Display
  Advertising}.
\newblock \bibinfo{journal}{\emph{CHINESE JOURNAL OF COMPUTERS}}
  \bibinfo{volume}{43}, \bibinfo{number}{10} (\bibinfo{year}{2020}),
  \bibinfo{pages}{1810--1841}.
\newblock


\bibitem[Liu et~al\mbox{.}(2014a)]%
        {liu2014measurement}
\bibfield{author}{\bibinfo{person}{Yabing Liu}, \bibinfo{person}{Chloe
  Kliman-Silver}, \bibinfo{person}{Robert Bell}, \bibinfo{person}{Balachander
  Krishnamurthy}, {and} \bibinfo{person}{Alan Mislove}.}
  \bibinfo{year}{2014}\natexlab{a}.
\newblock \showarticletitle{Measurement and analysis of osn ad auctions}. In
  \bibinfo{booktitle}{\emph{COSN}}. \bibinfo{pages}{139--150}.
\newblock


\bibitem[Ma et~al\mbox{.}(2018)]%
        {ma2018entire}
\bibfield{author}{\bibinfo{person}{Xiao Ma}, \bibinfo{person}{Liqin Zhao},
  \bibinfo{person}{Guan Huang}, \bibinfo{person}{Zhi Wang},
  \bibinfo{person}{Zelin Hu}, \bibinfo{person}{Xiaoqiang Zhu}, {and}
  \bibinfo{person}{Kun Gai}.} \bibinfo{year}{2018}\natexlab{}.
\newblock \showarticletitle{Entire space multi-task model: An effective
  approach for estimating post-click conversion rate}. In
  \bibinfo{booktitle}{\emph{SIGIR}}. \bibinfo{pages}{1137--1140}.
\newblock


\bibitem[Marlin et~al\mbox{.}(2012)]%
        {marlin2012collaborative}
\bibfield{author}{\bibinfo{person}{Benjamin Marlin}, \bibinfo{person}{Richard~S
  Zemel}, \bibinfo{person}{Sam Roweis}, {and} \bibinfo{person}{Malcolm
  Slaney}.} \bibinfo{year}{2012}\natexlab{}.
\newblock \showarticletitle{Collaborative filtering and the missing at random
  assumption}.
\newblock \bibinfo{journal}{\emph{arXiv:1206.5267}} (\bibinfo{year}{2012}).
\newblock


\bibitem[Mehrabi et~al\mbox{.}(2021)]%
        {mehrabi2021survey}
\bibfield{author}{\bibinfo{person}{Ninareh Mehrabi}, \bibinfo{person}{Fred
  Morstatter}, \bibinfo{person}{Nripsuta Saxena}, \bibinfo{person}{Kristina
  Lerman}, {and} \bibinfo{person}{Aram Galstyan}.}
  \bibinfo{year}{2021}\natexlab{}.
\newblock \showarticletitle{A survey on bias and fairness in machine learning}.
\newblock \bibinfo{journal}{\emph{ACM Computing Surveys (CSUR)}}
  \bibinfo{volume}{54}, \bibinfo{number}{6} (\bibinfo{year}{2021}),
  \bibinfo{pages}{1--35}.
\newblock


\bibitem[Mladenow et~al\mbox{.}(2015)]%
        {mladenow2015online}
\bibfield{author}{\bibinfo{person}{Andreas Mladenow},
  \bibinfo{person}{Niina~Maarit Novak}, {and} \bibinfo{person}{Christine
  Strauss}.} \bibinfo{year}{2015}\natexlab{}.
\newblock \showarticletitle{Online ad-fraud in search engine advertising
  campaigns}. In \bibinfo{booktitle}{\emph{ICT-EurAsia}}.
  \bibinfo{pages}{109--118}.
\newblock


\bibitem[Mungamuru et~al\mbox{.}(2008)]%
        {mungamuru2008should}
\bibfield{author}{\bibinfo{person}{Bob Mungamuru}, \bibinfo{person}{Stephen
  Weis}, {and} \bibinfo{person}{Hector Garcia-Molina}.}
  \bibinfo{year}{2008}\natexlab{}.
\newblock \bibinfo{booktitle}{\emph{Should ad networks bother fighting click
  fraud?(yes, they should.)}}.
\newblock \bibinfo{type}{{T}echnical {R}eport}.
  \bibinfo{institution}{Stanford}.
\newblock


\bibitem[Myerson(1981)]%
        {myerson1981optimal}
\bibfield{author}{\bibinfo{person}{Roger~B Myerson}.}
  \bibinfo{year}{1981}\natexlab{}.
\newblock \showarticletitle{Optimal auction design}.
\newblock \bibinfo{journal}{\emph{Mathematics of operations research}}
  \bibinfo{volume}{6}, \bibinfo{number}{1} (\bibinfo{year}{1981}),
  \bibinfo{pages}{58--73}.
\newblock


\bibitem[Narayanan(2018)]%
        {narayanan2018translation}
\bibfield{author}{\bibinfo{person}{Arvind Narayanan}.}
  \bibinfo{year}{2018}\natexlab{}.
\newblock \showarticletitle{Translation tutorial: 21 fairness definitions and
  their politics}. In \bibinfo{booktitle}{\emph{FAT}},
  Vol.~\bibinfo{volume}{1170}. \bibinfo{pages}{3}.
\newblock


\bibitem[Nasr and Tschantz(2020)]%
        {nasr2020bidding}
\bibfield{author}{\bibinfo{person}{Milad Nasr} {and}
  \bibinfo{person}{Michael~Carl Tschantz}.} \bibinfo{year}{2020}\natexlab{}.
\newblock \showarticletitle{Bidding strategies with gender nondiscrimination
  constraints for online ad auctions}. In \bibinfo{booktitle}{\emph{FAT}}.
  \bibinfo{pages}{337--347}.
\newblock


\bibitem[Nithyanand et~al\mbox{.}(2016)]%
        {nithyanand2016adblocking}
\bibfield{author}{\bibinfo{person}{Rishab Nithyanand},
  \bibinfo{person}{Sheharbano Khattak}, \bibinfo{person}{Mobin Javed},
  \bibinfo{person}{Narseo Vallina-Rodriguez}, \bibinfo{person}{Marjan
  Falahrastegar}, \bibinfo{person}{Julia~E Powles}, \bibinfo{person}{Emiliano
  De~Cristofaro}, \bibinfo{person}{Hamed Haddadi}, {and}
  \bibinfo{person}{Steven~J Murdoch}.} \bibinfo{year}{2016}\natexlab{}.
\newblock \showarticletitle{Adblocking and counter blocking: A slice of the
  arms race}. In \bibinfo{booktitle}{\emph{FOCI}}. \bibinfo{pages}{1--7}.
\newblock


\bibitem[O'Brien et~al\mbox{.}(2021)]%
        {o2021analysis}
\bibfield{author}{\bibinfo{person}{Conor O'Brien}, \bibinfo{person}{Kin~Sum
  Liu}, \bibinfo{person}{James Neufeld}, \bibinfo{person}{Rafael Barreto},
  {and} \bibinfo{person}{Jonathan~J Hunt}.} \bibinfo{year}{2021}\natexlab{}.
\newblock \showarticletitle{An Analysis Of Entire Space Multi-Task Models For
  Post-Click Conversion Prediction}. In \bibinfo{booktitle}{\emph{Recsys}}.
  \bibinfo{pages}{613--619}.
\newblock


\bibitem[Oentaryo et~al\mbox{.}(2014)]%
        {oentaryo2014detecting}
\bibfield{author}{\bibinfo{person}{Richard Oentaryo}, \bibinfo{person}{Ee-Peng
  Lim}, \bibinfo{person}{Michael Finegold}, \bibinfo{person}{David Lo},
  \bibinfo{person}{Feida Zhu}, \bibinfo{person}{Clifton Phua},
  \bibinfo{person}{Eng-Yeow Cheu}, \bibinfo{person}{Ghim-Eng Yap},
  \bibinfo{person}{Kelvin Sim}, \bibinfo{person}{Minh~Nhut Nguyen},
  {et~al\mbox{.}}} \bibinfo{year}{2014}\natexlab{}.
\newblock \showarticletitle{Detecting click fraud in online advertising: a data
  mining approach}.
\newblock \bibinfo{journal}{\emph{The Journal of Machine Learning Research}}
  \bibinfo{volume}{15}, \bibinfo{number}{1} (\bibinfo{year}{2014}),
  \bibinfo{pages}{99--140}.
\newblock


\bibitem[Oger et~al\mbox{.}(2015)]%
        {oger2015privacy}
\bibfield{author}{\bibinfo{person}{Mustafa Oger}, \bibinfo{person}{Isa Olmez},
  \bibinfo{person}{Erinc Inci}, \bibinfo{person}{Serkan K{\"u}c{\"u}kbay},
  {and} \bibinfo{person}{Fatih Emekci}.} \bibinfo{year}{2015}\natexlab{}.
\newblock \showarticletitle{Privacy preserving secure online advertising}.
\newblock \bibinfo{journal}{\emph{Procedia-Social and Behavioral Sciences}}
  \bibinfo{volume}{195} (\bibinfo{year}{2015}), \bibinfo{pages}{1840--1845}.
\newblock


\bibitem[Orr et~al\mbox{.}(2012)]%
        {orr2012approach}
\bibfield{author}{\bibinfo{person}{Caitlin~R Orr}, \bibinfo{person}{Arun
  Chauhan}, \bibinfo{person}{Minaxi Gupta}, \bibinfo{person}{Christopher~J
  Frisz}, {and} \bibinfo{person}{Christopher~W Dunn}.}
  \bibinfo{year}{2012}\natexlab{}.
\newblock \showarticletitle{An approach for identifying JavaScript-loaded
  advertisements through static program analysis}. In
  \bibinfo{booktitle}{\emph{WPES}}. \bibinfo{pages}{1--12}.
\newblock


\bibitem[Pan et~al\mbox{.}(2020)]%
        {pan2020bid}
\bibfield{author}{\bibinfo{person}{Shengjun Pan}, \bibinfo{person}{Brendan
  Kitts}, \bibinfo{person}{Tian Zhou}, \bibinfo{person}{Hao He},
  \bibinfo{person}{Bharatbhushan Shetty}, \bibinfo{person}{Aaron Flores},
  \bibinfo{person}{Djordje Gligorijevic}, \bibinfo{person}{Junwei Pan},
  \bibinfo{person}{Tingyu Mao}, \bibinfo{person}{San Gultekin},
  {et~al\mbox{.}}} \bibinfo{year}{2020}\natexlab{}.
\newblock \showarticletitle{Bid Shading by Win-Rate Estimation and Surplus
  Maximization}.
\newblock \bibinfo{journal}{\emph{arXiv:2009.09259}} (\bibinfo{year}{2020}).
\newblock


\bibitem[Patro et~al\mbox{.}(2020)]%
        {patro2020fairrec}
\bibfield{author}{\bibinfo{person}{Gourab~K Patro}, \bibinfo{person}{Arpita
  Biswas}, \bibinfo{person}{Niloy Ganguly}, \bibinfo{person}{Krishna~P
  Gummadi}, {and} \bibinfo{person}{Abhijnan Chakraborty}.}
  \bibinfo{year}{2020}\natexlab{}.
\newblock \showarticletitle{Fairrec: Two-sided fairness for personalized
  recommendations in two-sided platforms}. In \bibinfo{booktitle}{\emph{WWW}}.
  \bibinfo{pages}{1194--1204}.
\newblock


\bibitem[Pearce et~al\mbox{.}(2014)]%
        {pearce2014characterizing}
\bibfield{author}{\bibinfo{person}{Paul Pearce}, \bibinfo{person}{Vacha Dave},
  \bibinfo{person}{Chris Grier}, \bibinfo{person}{Kirill Levchenko},
  \bibinfo{person}{Saikat Guha}, \bibinfo{person}{Damon McCoy},
  \bibinfo{person}{Vern Paxson}, \bibinfo{person}{Stefan Savage}, {and}
  \bibinfo{person}{Geoffrey~M Voelker}.} \bibinfo{year}{2014}\natexlab{}.
\newblock \showarticletitle{Characterizing large-scale click fraud in
  zeroaccess}. In \bibinfo{booktitle}{\emph{CCS}}. \bibinfo{pages}{141--152}.
\newblock


\bibitem[Pearce et~al\mbox{.}(2012)]%
        {pearce2012addroid}
\bibfield{author}{\bibinfo{person}{Paul Pearce},
  \bibinfo{person}{Adrienne~Porter Felt}, \bibinfo{person}{Gabriel Nunez},
  {and} \bibinfo{person}{David Wagner}.} \bibinfo{year}{2012}\natexlab{}.
\newblock \showarticletitle{Addroid: Privilege separation for applications and
  advertisers in android}. In \bibinfo{booktitle}{\emph{ASIACCS}}.
  \bibinfo{pages}{71--72}.
\newblock


\bibitem[Peri et~al\mbox{.}(2021)]%
        {peri2021preferencenet}
\bibfield{author}{\bibinfo{person}{Neehar Peri}, \bibinfo{person}{Michael
  Curry}, \bibinfo{person}{Samuel Dooley}, {and} \bibinfo{person}{John
  Dickerson}.} \bibinfo{year}{2021}\natexlab{}.
\newblock \showarticletitle{Preferencenet: Encoding human preferences in
  auction design with deep learning}.
\newblock  (\bibinfo{year}{2021}), \bibinfo{pages}{17532--17542}.
\newblock


\bibitem[Pooranian et~al\mbox{.}(2021)]%
        {pooranian2021online}
\bibfield{author}{\bibinfo{person}{Zahra Pooranian}, \bibinfo{person}{Mauro
  Conti}, \bibinfo{person}{Hamed Haddadi}, {and} \bibinfo{person}{Rahim
  Tafazolli}.} \bibinfo{year}{2021}\natexlab{}.
\newblock \showarticletitle{Online advertising security: Issues, taxonomy, and
  future directions}.
\newblock \bibinfo{journal}{\emph{IEEE Communications Surveys \& Tutorials}}
  \bibinfo{volume}{3}, \bibinfo{number}{4} (\bibinfo{year}{2021}),
  \bibinfo{pages}{2494, 2524}.
\newblock


\bibitem[Poornachandran et~al\mbox{.}(2017)]%
        {poornachandran2017demalvertising}
\bibfield{author}{\bibinfo{person}{Prabaharan Poornachandran},
  \bibinfo{person}{N Balagopal}, \bibinfo{person}{Soumajit Pal},
  \bibinfo{person}{Aravind Ashok}, \bibinfo{person}{Prem Sankar}, {and}
  \bibinfo{person}{Manu~R Krishnan}.} \bibinfo{year}{2017}\natexlab{}.
\newblock \showarticletitle{Demalvertising: A kernel approach for detecting
  malwares in advertising networks}. In \bibinfo{booktitle}{\emph{ICIC2}}.
  \bibinfo{pages}{215--224}.
\newblock


\bibitem[Ray et~al\mbox{.}(2017)]%
        {ray2017ad}
\bibfield{author}{\bibinfo{person}{Abhishek Ray}, \bibinfo{person}{Hossein
  Ghasemkhani}, {and} \bibinfo{person}{Karthik~N Kannan}.}
  \bibinfo{year}{2017}\natexlab{}.
\newblock \showarticletitle{Ad-blockers, advertisers, and internet: The
  economic implications of ad-blocker platforms}. In
  \bibinfo{booktitle}{\emph{ICIS}}. \bibinfo{pages}{140--152}.
\newblock


\bibitem[Reis et~al\mbox{.}(2008)]%
        {reis2008detecting}
\bibfield{author}{\bibinfo{person}{Charles Reis}, \bibinfo{person}{Steven~D
  Gribble}, \bibinfo{person}{Tadayoshi Kohno}, {and}
  \bibinfo{person}{Nicholas~C Weaver}.} \bibinfo{year}{2008}\natexlab{}.
\newblock \showarticletitle{Detecting In-Flight Page Changes with Web
  Tripwires}. In \bibinfo{booktitle}{\emph{NSDI}}, Vol.~\bibinfo{volume}{8}.
  \bibinfo{pages}{31--44}.
\newblock


\bibitem[Ren et~al\mbox{.}(2019)]%
        {ren2019deep}
\bibfield{author}{\bibinfo{person}{Kan Ren}, \bibinfo{person}{Jiarui Qin},
  \bibinfo{person}{Lei Zheng}, \bibinfo{person}{Zhengyu Yang},
  \bibinfo{person}{Weinan Zhang}, {and} \bibinfo{person}{Yong Yu}.}
  \bibinfo{year}{2019}\natexlab{}.
\newblock \showarticletitle{Deep landscape forecasting for real-time bidding
  advertising}. In \bibinfo{booktitle}{\emph{KDD}}. \bibinfo{pages}{363--372}.
\newblock


\bibitem[Rubicon.(2018)]%
        {rubicon2018}
Rubicon. \bibinfo{year}{2018}\natexlab{}.
\newblock \bibinfo{booktitle}{\emph{Principles for a Better Programmatic
  Marketplace, Open Letter from Rubicon, SpotX, OpenX, Pubmatic, Telaris, and
  Sovr}}.
\newblock
\urldef\tempurl%
\url{https://rubiconproject.com/insights/thought-leadership/principles-betterprogrammatic-marketplace-open-letter-advertisers-publishers/}
\showURL{%
Retrieved Jun 01, 2022 from \tempurl}


\bibitem[S. Sluis.(2017)]%
        {ssluis2017}
S. Sluis. \bibinfo{year}{2017}\natexlab{}.
\newblock \bibinfo{booktitle}{\emph{Explainer: More On The Widespread Fee
  Practice Behind The Guardian’s Lawsuit Vs. Rubicon Project}}.
\newblock
\urldef\tempurl%
\url{https://adexchanger.com/adexchange-news/explainer-widespread-fee-practice-behind-guardians-lawsuitvs-rubicon-project/}
\showURL{%
Retrieved Jun 01, 2022 from \tempurl}


\bibitem[Saez-Trumper et~al\mbox{.}(2014)]%
        {saez2014beyond}
\bibfield{author}{\bibinfo{person}{Diego Saez-Trumper}, \bibinfo{person}{Yabing
  Liu}, \bibinfo{person}{Ricardo Baeza-Yates}, \bibinfo{person}{Balachander
  Krishnamurthy}, {and} \bibinfo{person}{Alan Mislove}.}
  \bibinfo{year}{2014}\natexlab{}.
\newblock \showarticletitle{Beyond cpm and cpc: Determining the value of users
  on osns}. In \bibinfo{booktitle}{\emph{COSN}}. \bibinfo{pages}{161--168}.
\newblock


\bibitem[Sakib and Huang(2015)]%
        {sakib2015automated}
\bibfield{author}{\bibinfo{person}{Muhammad~N Sakib} {and}
  \bibinfo{person}{Chin-Tser Huang}.} \bibinfo{year}{2015}\natexlab{}.
\newblock \showarticletitle{Automated collection and analysis of malware
  disseminated via online advertising}. In
  \bibinfo{booktitle}{\emph{TRUSTCOM}}. \bibinfo{pages}{1411--1416}.
\newblock


\bibitem[Sarah Sluis.(2017)]%
        {sarah2017}
Sarah Sluis. \bibinfo{year}{2017}\natexlab{}.
\newblock \bibinfo{booktitle}{\emph{Big Changes Coming To Auctions, As
  Exchanges Roll The Dice On First-Price}}.
\newblock
\urldef\tempurl%
\url{https://www.adexchanger.com/platforms/big-changes-coming-auctions-exchanges-roll-dice-first-price/}
\showURL{%
Retrieved Jun 01, 2022 from \tempurl}


\bibitem[Saxena et~al\mbox{.}(2019)]%
        {saxena2019fairness}
\bibfield{author}{\bibinfo{person}{Nripsuta~Ani Saxena}, \bibinfo{person}{Karen
  Huang}, \bibinfo{person}{Evan DeFilippis}, \bibinfo{person}{Goran Radanovic},
  \bibinfo{person}{David~C Parkes}, {and} \bibinfo{person}{Yang Liu}.}
  \bibinfo{year}{2019}\natexlab{}.
\newblock \showarticletitle{How do fairness definitions fare? Examining public
  attitudes towards algorithmic definitions of fairness}. In
  \bibinfo{booktitle}{\emph{AIES}}. \bibinfo{pages}{99--106}.
\newblock


\bibitem[Shah et~al\mbox{.}(2019)]%
        {shah2019predictive}
\bibfield{author}{\bibinfo{person}{Deven Shah}, \bibinfo{person}{H~Andrew
  Schwartz}, {and} \bibinfo{person}{Dirk Hovy}.}
  \bibinfo{year}{2019}\natexlab{}.
\newblock \showarticletitle{Predictive biases in natural language processing
  models: A conceptual framework and overview}.
\newblock \bibinfo{journal}{\emph{arXiv:1912.11078}} (\bibinfo{year}{2019}).
\newblock


\bibitem[Shekhar et~al\mbox{.}(2012)]%
        {shekhar2012adsplit}
\bibfield{author}{\bibinfo{person}{Shashi Shekhar}, \bibinfo{person}{Michael
  Dietz}, {and} \bibinfo{person}{Dan~S Wallach}.}
  \bibinfo{year}{2012}\natexlab{}.
\newblock \showarticletitle{AdSplit: Separating Smartphone Advertising from
  Applications}. In \bibinfo{booktitle}{\emph{USENIX Security}}.
  \bibinfo{pages}{553--567}.
\newblock


\bibitem[Shi et~al\mbox{.}(2020)]%
        {shi2020clickguard}
\bibfield{author}{\bibinfo{person}{Congcong Shi}, \bibinfo{person}{Rui Song},
  \bibinfo{person}{Xinyu Qi}, \bibinfo{person}{Yubo Song}, \bibinfo{person}{Bin
  Xiao}, {and} \bibinfo{person}{Sanglu Lu}.} \bibinfo{year}{2020}\natexlab{}.
\newblock \showarticletitle{ClickGuard: Exposing hidden click fraud via mobile
  sensor side-channel analysis}. In \bibinfo{booktitle}{\emph{ICC}}. IEEE,
  \bibinfo{pages}{1--6}.
\newblock


\bibitem[Singh et~al\mbox{.}(2012)]%
        {singh2012practical}
\bibfield{author}{\bibinfo{person}{Kapil Singh}, \bibinfo{person}{Helen~J
  Wang}, \bibinfo{person}{Alexander Moshchuk}, \bibinfo{person}{Collin
  Jackson}, {and} \bibinfo{person}{Wenke Lee}.}
  \bibinfo{year}{2012}\natexlab{}.
\newblock \showarticletitle{Practical end-to-end web content integrity}. In
  \bibinfo{booktitle}{\emph{WWW}}. \bibinfo{pages}{659--668}.
\newblock


\bibitem[Smuha(2019)]%
        {smuha2019eu}
\bibfield{author}{\bibinfo{person}{Nathalie~A Smuha}.}
  \bibinfo{year}{2019}\natexlab{}.
\newblock \showarticletitle{The EU approach to ethics guidelines for
  trustworthy artificial intelligence}.
\newblock \bibinfo{journal}{\emph{Computer Law Review International}}
  \bibinfo{volume}{20}, \bibinfo{number}{4} (\bibinfo{year}{2019}),
  \bibinfo{pages}{97--106}.
\newblock


\bibitem[Sobel and Satish(2012)]%
        {sobel2012methods}
\bibfield{author}{\bibinfo{person}{William~E Sobel} {and}
  \bibinfo{person}{Sourabh Satish}.} \bibinfo{year}{2012}\natexlab{}.
\newblock \bibinfo{title}{Methods and systems for detecting man-in-the-browser
  attacks}.
\newblock
\newblock
\newblock
\shownote{US Patent 8,225,401}.


\bibitem[Stone-Gross et~al\mbox{.}(2011)]%
        {stone2011understanding}
\bibfield{author}{\bibinfo{person}{Brett Stone-Gross}, \bibinfo{person}{Ryan
  Stevens}, \bibinfo{person}{Apostolis Zarras}, \bibinfo{person}{Richard
  Kemmerer}, \bibinfo{person}{Chris Kruegel}, {and} \bibinfo{person}{Giovanni
  Vigna}.} \bibinfo{year}{2011}\natexlab{}.
\newblock \showarticletitle{Understanding fraudulent activities in online ad
  exchanges}. In \bibinfo{booktitle}{\emph{IMC}}. \bibinfo{pages}{279--294}.
\newblock


\bibitem[Thiebes et~al\mbox{.}(2021)]%
        {thiebes2021trustworthy}
\bibfield{author}{\bibinfo{person}{Scott Thiebes}, \bibinfo{person}{Sebastian
  Lins}, {and} \bibinfo{person}{Ali Sunyaev}.} \bibinfo{year}{2021}\natexlab{}.
\newblock \showarticletitle{Trustworthy artificial intelligence}.
\newblock \bibinfo{journal}{\emph{Electronic Markets}} \bibinfo{volume}{31},
  \bibinfo{number}{2} (\bibinfo{year}{2021}), \bibinfo{pages}{447--464}.
\newblock


\bibitem[Tian et~al\mbox{.}(2015)]%
        {tian2015crowd}
\bibfield{author}{\bibinfo{person}{Tian Tian}, \bibinfo{person}{Jun Zhu},
  \bibinfo{person}{Fen Xia}, \bibinfo{person}{Xin Zhuang}, {and}
  \bibinfo{person}{Tong Zhang}.} \bibinfo{year}{2015}\natexlab{}.
\newblock \showarticletitle{Crowd fraud detection in internet advertising}. In
  \bibinfo{booktitle}{\emph{WWW}}. \bibinfo{pages}{1100--1110}.
\newblock


\bibitem[Turkel(2022)]%
        {turkel2022regulating}
\bibfield{author}{\bibinfo{person}{Eray Turkel}.}
  \bibinfo{year}{2022}\natexlab{}.
\newblock \showarticletitle{Regulating Online Political Advertising}. In
  \bibinfo{booktitle}{\emph{WWW}}. \bibinfo{pages}{3584--3593}.
\newblock


\bibitem[Vermeren, I.(2015)]%
        {vermeren2015}
Vermeren, I. \bibinfo{year}{2015}\natexlab{}.
\newblock \bibinfo{booktitle}{\emph{Men vs. Women: Who Is More Active on Social
  Media?}}
\newblock
\urldef\tempurl%
\url{http://bit.ly/ 2Xvy6xY}
\showURL{%
Retrieved Jun 01, 2022 from \tempurl}


\bibitem[Vratonjic et~al\mbox{.}(2010a)]%
        {vratonjic2010integrity}
\bibfield{author}{\bibinfo{person}{Nevena Vratonjic}, \bibinfo{person}{Julien
  Freudiger}, {and} \bibinfo{person}{Jean-Pierre Hubaux}.}
  \bibinfo{year}{2010}\natexlab{a}.
\newblock \showarticletitle{Integrity of the web content: The case of online
  advertising}. In \bibinfo{booktitle}{\emph{CollSec}}. \bibinfo{pages}{1--12}.
\newblock


\bibitem[Vratonjic et~al\mbox{.}(2010b)]%
        {vratonjic2010isps}
\bibfield{author}{\bibinfo{person}{Nevena Vratonjic},
  \bibinfo{person}{Mohammad~Hossein Manshaei}, \bibinfo{person}{Maxim Raya},
  {and} \bibinfo{person}{Jean-Pierre Hubaux}.}
  \bibinfo{year}{2010}\natexlab{b}.
\newblock \showarticletitle{ISPs and ad networks against botnet ad fraud}. In
  \bibinfo{booktitle}{\emph{GameSec}}. \bibinfo{pages}{149--167}.
\newblock


\bibitem[Wang et~al\mbox{.}(2017)]%
        {wang2017display}
\bibfield{author}{\bibinfo{person}{Jun Wang}, \bibinfo{person}{Weinan Zhang},
  \bibinfo{person}{Shuai Yuan}, {et~al\mbox{.}}}
  \bibinfo{year}{2017}\natexlab{}.
\newblock \showarticletitle{Display advertising with real-time bidding (RTB)
  and behavioural targeting}.
\newblock \bibinfo{journal}{\emph{Foundations and Trends{\textregistered} in
  Information Retrieval}} \bibinfo{volume}{11}, \bibinfo{number}{4-5}
  (\bibinfo{year}{2017}), \bibinfo{pages}{297--435}.
\newblock


\bibitem[Wang et~al\mbox{.}(2022b)]%
        {wang2022kaplan}
\bibfield{author}{\bibinfo{person}{Tengyun Wang}, \bibinfo{person}{Haizhi
  Yang}, \bibinfo{person}{Siyu Jiang}, \bibinfo{person}{Yueyue Shi},
  \bibinfo{person}{Qianyu Li}, \bibinfo{person}{Xiaoli Tang},
  \bibinfo{person}{Han Yu}, {and} \bibinfo{person}{Hengjie Song}.}
  \bibinfo{year}{2022}\natexlab{b}.
\newblock \showarticletitle{Kaplan--Meier Markov network: Learning the
  distribution of market price by censored data in online advertising}.
\newblock \bibinfo{journal}{\emph{Knowledge-Based Systems}}
  \bibinfo{volume}{251} (\bibinfo{year}{2022}), \bibinfo{pages}{109248}.
\newblock
\showISSN{0950-7051}


\bibitem[Wang et~al\mbox{.}(2022a)]%
        {wang2022survey}
\bibfield{author}{\bibinfo{person}{Yifan Wang}, \bibinfo{person}{Weizhi Ma},
  \bibinfo{person}{Min Zhang}, \bibinfo{person}{Yiqun Liu}, {and}
  \bibinfo{person}{Shaoping Ma}.} \bibinfo{year}{2022}\natexlab{a}.
\newblock \showarticletitle{A Survey on the Fairness of Recommender Systems}.
\newblock \bibinfo{journal}{\emph{arXiv:2206.03761}} (\bibinfo{year}{2022}).
\newblock


\bibitem[Wang et~al\mbox{.}(2020)]%
        {wang2020delayed}
\bibfield{author}{\bibinfo{person}{Yanshi Wang}, \bibinfo{person}{Jie Zhang},
  \bibinfo{person}{Qing Da}, {and} \bibinfo{person}{Anxiang Zeng}.}
  \bibinfo{year}{2020}\natexlab{}.
\newblock \showarticletitle{Delayed feedback modeling for the entire space
  conversion rate prediction}.
\newblock \bibinfo{journal}{\emph{arXiv:2011.11826}} (\bibinfo{year}{2020}).
\newblock


\bibitem[Wei et~al\mbox{.}(2021)]%
        {wei2021autoheri}
\bibfield{author}{\bibinfo{person}{Penghui Wei}, \bibinfo{person}{Weimin
  Zhang}, \bibinfo{person}{Zixuan Xu}, \bibinfo{person}{Shaoguo Liu},
  \bibinfo{person}{Kuang-chih Lee}, {and} \bibinfo{person}{Bo Zheng}.}
  \bibinfo{year}{2021}\natexlab{}.
\newblock \showarticletitle{AutoHERI: Automated Hierarchical Representation
  Integration for Post-Click Conversion Rate Estimation}. In
  \bibinfo{booktitle}{\emph{CIKM}}. \bibinfo{pages}{3528--3532}.
\newblock


\bibitem[Wen et~al\mbox{.}(2021)]%
        {wen2021hierarchically}
\bibfield{author}{\bibinfo{person}{Hong Wen}, \bibinfo{person}{Jing Zhang},
  \bibinfo{person}{Fuyu Lv}, \bibinfo{person}{Wentian Bao},
  \bibinfo{person}{Tianyi Wang}, {and} \bibinfo{person}{Zulong Chen}.}
  \bibinfo{year}{2021}\natexlab{}.
\newblock \showarticletitle{Hierarchically modeling micro and macro behaviors
  via multi-task learning for conversion rate prediction}. In
  \bibinfo{booktitle}{\emph{SIGIR}}. \bibinfo{pages}{2187--2191}.
\newblock


\bibitem[Wen et~al\mbox{.}(2020)]%
        {wen2020entire}
\bibfield{author}{\bibinfo{person}{Hong Wen}, \bibinfo{person}{Jing Zhang},
  \bibinfo{person}{Yuan Wang}, \bibinfo{person}{Fuyu Lv},
  \bibinfo{person}{Wentian Bao}, \bibinfo{person}{Quan Lin}, {and}
  \bibinfo{person}{Keping Yang}.} \bibinfo{year}{2020}\natexlab{}.
\newblock \showarticletitle{Entire space multi-task modeling via post-click
  behavior decomposition for conversion rate prediction}. In
  \bibinfo{booktitle}{\emph{SIGIR}}. \bibinfo{pages}{2377--2386}.
\newblock


\bibitem[Wolpert and Macready(1997)]%
        {wolpert1997no}
\bibfield{author}{\bibinfo{person}{David~H Wolpert} {and}
  \bibinfo{person}{William~G Macready}.} \bibinfo{year}{1997}\natexlab{}.
\newblock \showarticletitle{No free lunch theorems for optimization}.
\newblock \bibinfo{journal}{\emph{IEEE transactions on evolutionary
  computation}} \bibinfo{volume}{1}, \bibinfo{number}{1}
  (\bibinfo{year}{1997}), \bibinfo{pages}{67--82}.
\newblock


\bibitem[Wu et~al\mbox{.}(2018)]%
        {wu2018deep}
\bibfield{author}{\bibinfo{person}{Wush Wu}, \bibinfo{person}{Mi-Yen Yeh},
  {and} \bibinfo{person}{Ming-Syan Chen}.} \bibinfo{year}{2018}\natexlab{}.
\newblock \showarticletitle{Deep censored learning of the winning price in the
  real time bidding}. In \bibinfo{booktitle}{\emph{KDD}}.
  \bibinfo{pages}{2526--2535}.
\newblock


\bibitem[Wu et~al\mbox{.}(2015)]%
        {wu2015predicting}
\bibfield{author}{\bibinfo{person}{Wush Chi-Hsuan Wu}, \bibinfo{person}{Mi-Yen
  Yeh}, {and} \bibinfo{person}{Ming-Syan Chen}.}
  \bibinfo{year}{2015}\natexlab{}.
\newblock \showarticletitle{Predicting winning price in real time bidding with
  censored data}. In \bibinfo{booktitle}{\emph{KDD}}.
  \bibinfo{pages}{1305--1314}.
\newblock


\bibitem[Wu et~al\mbox{.}(2019)]%
        {wu2019counterfactual}
\bibfield{author}{\bibinfo{person}{Yongkai Wu}, \bibinfo{person}{Lu Zhang},
  {and} \bibinfo{person}{Xintao Wu}.} \bibinfo{year}{2019}\natexlab{}.
\newblock \showarticletitle{Counterfactual fairness: Unidentification, bound
  and algorithm}. In \bibinfo{booktitle}{\emph{IJCAI}}.
  \bibinfo{pages}{1438--1444}.
\newblock


\bibitem[Xu et~al\mbox{.}(2014)]%
        {xu2014click}
\bibfield{author}{\bibinfo{person}{Haitao Xu}, \bibinfo{person}{Daiping Liu},
  \bibinfo{person}{Aaron Koehl}, \bibinfo{person}{Haining Wang}, {and}
  \bibinfo{person}{Angelos Stavrou}.} \bibinfo{year}{2014}\natexlab{}.
\newblock \showarticletitle{Click fraud detection on the advertiser side}. In
  \bibinfo{booktitle}{\emph{ESORICS}}. \bibinfo{pages}{419--438}.
\newblock


\bibitem[Yang et~al\mbox{.}(2021)]%
        {yang2021multi}
\bibfield{author}{\bibinfo{person}{Haizhi Yang}, \bibinfo{person}{Tengyun
  Wang}, \bibinfo{person}{Xiaoli Tang}, \bibinfo{person}{Qianyu Li},
  \bibinfo{person}{Yueyue Shi}, \bibinfo{person}{Siyu Jiang},
  \bibinfo{person}{Han Yu}, {and} \bibinfo{person}{Hengjie Song}.}
  \bibinfo{year}{2021}\natexlab{}.
\newblock \showarticletitle{Multi-task Learning for Bias-Free Joint CTR
  Prediction and Market Price Modeling in Online Advertising}. In
  \bibinfo{booktitle}{\emph{CIKM}}. \bibinfo{pages}{2291--2300}.
\newblock


\bibitem[Zadrozny(2004)]%
        {zadrozny2004learning}
\bibfield{author}{\bibinfo{person}{Bianca Zadrozny}.}
  \bibinfo{year}{2004}\natexlab{}.
\newblock \showarticletitle{Learning and evaluating classifiers under sample
  selection bias}. In \bibinfo{booktitle}{\emph{ICML}}. \bibinfo{pages}{114}.
\newblock


\bibitem[Zhang et~al\mbox{.}(2020)]%
        {zhang2020large}
\bibfield{author}{\bibinfo{person}{Wenhao Zhang}, \bibinfo{person}{Wentian
  Bao}, \bibinfo{person}{Xiao-Yang Liu}, \bibinfo{person}{Keping Yang},
  \bibinfo{person}{Quan Lin}, \bibinfo{person}{Hong Wen}, {and}
  \bibinfo{person}{Ramin Ramezani}.} \bibinfo{year}{2020}\natexlab{}.
\newblock \showarticletitle{Large-scale causal approaches to debiasing
  post-click conversion rate estimation with multi-task learning}. In
  \bibinfo{booktitle}{\emph{WWW}}. \bibinfo{pages}{2775--2781}.
\newblock


\bibitem[Zhang et~al\mbox{.}(2016)]%
        {zhang2016bid}
\bibfield{author}{\bibinfo{person}{Weinan Zhang}, \bibinfo{person}{Tianxiong
  Zhou}, \bibinfo{person}{Jun Wang}, {and} \bibinfo{person}{Jian Xu}.}
  \bibinfo{year}{2016}\natexlab{}.
\newblock \showarticletitle{Bid-aware gradient descent for unbiased learning
  with censored data in display advertising}. In
  \bibinfo{booktitle}{\emph{KDD}}. \bibinfo{pages}{665--674}.
\newblock


\bibitem[Zhou et~al\mbox{.}(2021)]%
        {zhou2021efficient}
\bibfield{author}{\bibinfo{person}{Tian Zhou}, \bibinfo{person}{Hao He},
  \bibinfo{person}{Shengjun Pan}, \bibinfo{person}{Niklas Karlsson},
  \bibinfo{person}{Bharatbhushan Shetty}, \bibinfo{person}{Brendan Kitts},
  \bibinfo{person}{Djordje Gligorijevic}, \bibinfo{person}{San Gultekin},
  \bibinfo{person}{Tingyu Mao}, \bibinfo{person}{Junwei Pan}, {et~al\mbox{.}}}
  \bibinfo{year}{2021}\natexlab{}.
\newblock \showarticletitle{An efficient deep distribution network for bid
  shading in first-price auctions}. In \bibinfo{booktitle}{\emph{KDD}}.
  \bibinfo{pages}{3996--4004}.
\newblock


\bibitem[Zhu et~al\mbox{.}(2018)]%
        {zhu2018measuring}
\bibfield{author}{\bibinfo{person}{Shitong Zhu}, \bibinfo{person}{Xunchao Hu},
  \bibinfo{person}{Zhiyun Qian}, \bibinfo{person}{Zubair Shafiq}, {and}
  \bibinfo{person}{Heng Yin}.} \bibinfo{year}{2018}\natexlab{}.
\newblock \showarticletitle{Measuring and disrupting anti-adblockers using
  differential execution analysis}. In \bibinfo{booktitle}{\emph{NDSS}}.
  \bibinfo{pages}{1--15}.
\newblock


\bibitem[Zhu et~al\mbox{.}(2017a)]%
        {zhu2017gamma}
\bibfield{author}{\bibinfo{person}{Wen-Yuan Zhu}, \bibinfo{person}{Wen-Yueh
  Shih}, \bibinfo{person}{Ying-Hsuan Lee}, \bibinfo{person}{Wen-Chih Peng},
  {and} \bibinfo{person}{Jiun-Long Huang}.} \bibinfo{year}{2017}\natexlab{a}.
\newblock \showarticletitle{A gamma-based regression for winning price
  estimation in real-time bidding advertising}. In
  \bibinfo{booktitle}{\emph{Big Data}}. \bibinfo{pages}{1610--1619}.
\newblock


\bibitem[Zhu et~al\mbox{.}(2017b)]%
        {zhu2017fraud}
\bibfield{author}{\bibinfo{person}{Xingquan Zhu}, \bibinfo{person}{Haicheng
  Tao}, \bibinfo{person}{Zhiang Wu}, \bibinfo{person}{Jie Cao},
  \bibinfo{person}{Kristopher Kalish}, {and} \bibinfo{person}{Jeremy Kayne}.}
  \bibinfo{year}{2017}\natexlab{b}.
\newblock \bibinfo{booktitle}{\emph{Fraud prevention in online digital
  advertising}}.
\newblock \bibinfo{publisher}{Springer}.
\newblock


\end{thebibliography}
%\bibliography{cites_simple}

%%
%% If your work has an appendix, this is the place to put it.
% \appendix

\end{document}